\documentclass{article}

\usepackage{PRIMEarxiv}

\usepackage[utf8]{inputenc} 
\usepackage[T1]{fontenc}    
\usepackage{hyperref}       
\usepackage{url}            
\usepackage{booktabs}       
\usepackage{amsfonts}       
\usepackage{nicefrac}       
\usepackage{microtype}      
\usepackage{lipsum}
\usepackage{fancyhdr}       
\usepackage{graphicx}       
\graphicspath{{media/}}     

\title{Fast M\"obius and Zeta Transforms\thanks{This is paper is currently under review.}
}

\author{
  Tommaso Pegolotti\\
  Department of Computer Science \\
  ETH Zurich \\
  \texttt{tommaso.pegolotti@inf.ethz.ch} \\
   \And
  Bastian Seifert \\
  Department of Computer Science \\
  ETH Zurich \\
  \texttt{bastian.seifert@inf.ethz.ch} \\
  \And
  Markus P\"uschel \\
  Department of Computer Science \\
  ETH Zurich \\
  \texttt{pueschel@inf.ethz.ch} \\
}

\usepackage{lipsum}
\usepackage{amsfonts}
\usepackage{mathrsfs}
\usepackage{graphicx}
\usepackage{graphbox}
\usepackage{epstopdf}
\usepackage{amsmath}
\usepackage{amssymb}
\usepackage{caption}
\usepackage{cleveref}
\usepackage{subcaption}
\usepackage{algorithmic}
\usepackage{algorithm}
\ifpdf
  \DeclareGraphicsExtensions{.eps,.pdf,.png,.jpg}
\else
  \DeclareGraphicsExtensions{.eps}
\fi

\usepackage{blkarray} 
\usepackage{stackengine}

\usepackage{amsopn}

\newcommand{\R}{\mathbb{R}}
\newcommand{\Z}{\mathbb{Z}}

\newcommand{\E}{\mathcal{E}}
\newcommand{\Ered}{\mathcal{E}_\text{red}}
\newcommand{\Eclose}{\mathcal{E}_\text{close}}
\newcommand{\G}{\mathcal{G}}
\newcommand{\myP}{\mathcal{P}}

\newcommand{\A}{\mathcal{A}}

\newcommand{\Lc}{\mathcal{L}}
\newcommand{\Zc}{\mathcal{Z}}
\newcommand{\On}{\mathcal{O}}

\newcommand{\id}{\textit{id}}
\newcommand{\next}{\textit{next}}
\newcommand{\adj}{\textit{adj}}
\newcommand{\width}{\textit{width}}
\newcommand{\length}{\textit{length}}

\newcommand{\n}{n}
\newcommand{\q}{q}
\newcommand{\nivsize}{\textit{nnz}(N)}

\newcommand{\zeroel}{\cdot}
\newcommand{\zeroelph}{\phantom{-}\zeroel}

\newcommand{\niv}{\textit{niv}}

\newcommand{\mypar}[1]{{\bf #1.}}

\newcommand{\closure}[1]{\textit{out}(#1)}

\newcommand{\coord}[1]{\text{\bf #1}}

\newcommand{\partleq}{\preceq}
\newcommand{\partl}{\prec}
\newcommand{\npartleq}{\npreceq}

\newtheorem{theorem}{Theorem}
\newtheorem{corollary}{Corollary}
\newtheorem{definition}{Definition}

\newenvironment{psmallmatrix}
  {\left(\begin{smallmatrix}}
  {\end{smallmatrix}\right)}
  
\newenvironment{xsmallmatrix}[1]
  {\renewcommand\thickspace{\kern#1}\smallmatrix}
  {\endsmallmatrix}

\begin{document}
\maketitle

\begin{abstract}
M\"obius inversion of functions on partially ordered sets (posets) $\myP$ is a classical tool in combinatorics. For finite posets it consists of two, mutually inverse, linear transformations called zeta and M\"obius transform, respectively. In this paper we provide novel fast algorithms for both
that require $O(nk)$ time and space, where $n = |\myP|$ and $k$ is the width (length of longest antichain) of $\myP$, compared to $O(n^2)$ for a direct computation. Our approach assumes that $\myP$ is given as directed acyclic graph (DAG) $(\E, \myP)$. The algorithms are then constructed using a chain decomposition for a one time cost of $\On(|\E| + |\Ered| \cdot k)$, where $\Ered$ is the number of edges in the DAG's transitive reduction. We show benchmarks with implementations of all algorithms including parallelized versions. The results show that our algorithms enable M\"obius inversion on posets with millions of nodes in seconds if the defining DAGs are sufficiently sparse.
\end{abstract}

\section{Introduction}
\label{sec:intro}
The M\"obius inversion formula is a classical tool in combinatorics~\cite{rota,stanley_2011}. Given a finite, partially ordered set (poset) $(\myP, \partleq)$  and two functions $f, g : \myP \rightarrow \R$ it states that
\begin{equation}
\label{eq:inv-formula}
	g(x) = \sum_{y \partleq x} f(y)\quad\text{ if and only if }\quad f(x) = \sum_{y \partleq x} \mu(y,x)g(y).
\end{equation}
Here $\mu$ is the so-called M\"obius function that is recursively defined as $\mu(x,y) = - \smash{\sum\nolimits_{x \partleq z \partl y}}\mu(x,z)$ if $x \neq y$ and $1$ if $x = y$. The left and right equation in~\eqref{eq:inv-formula} define two linear mappings on the vector space of poset functions. They are inverses of each other and known as \emph{zeta} and \emph{M\"obius} transform, respectively. This paper is concerned with the efficient computation of these transforms on arbitrary finite posets.

We assume the poset is given as a directed acyclic connected graph (DAG) $\G=(\myP,\E)$ that defines the partial order as $x \partleq y$ if and only if $x$ is a successor of $y$ in $\G$. A naive implementation of the zeta transform would first (as precomputation) explicitly construct the transform matrix by computing the transitive closure $(\myP, \Eclose)$ of $\G$ in $O(|\myP|\cdot|\E|)$. The transform can then be computed in $\On(|\Eclose|) \subseteq \On(|\myP|^2)$ space and time. The inverse (M\"obius) transform can be computed with the same complexity by solving a linear system.


The time and memory requirement does not scale to large sizes. In this paper, we provide an algorithm that improves both to $\On(|\myP| \cdot k)$, where $k = \width(\myP)$ is the length of the longest antichain in $\G$. Our approach also requires a precomputation step that is needed only once for a given poset and that has the same complexity as in the naive solution. First, we compute an optimal chain decomposition of the DAG by solving a maximum matching problem in $\On(|\myP| \cdot |\E|)$~\cite{fulkerson}. Second, we use this decomposition to construct an associated map on $\myP$ in $\On(|\E| + |\Ered| \cdot k)$ that was proposed in \cite{klaus-simon} to answer reachability questions. $\Ered$ is the set of edges in the cover graph of the poset. Using this map, both transforms can be computed in in $\On(|\myP| \cdot k)$. We implemented the entire approach and demonstrate that it enables the computation of both transforms on sparse DAGs up to millions of elements on a standard current work station.


\mypar{Related work} To date fast algorithm for zeta and M\"obius transform have been developed only for special classes of posets, and in particular lattices, in which each two elements have a unique largest lower bound and unique smallest upper bound \cite{Graetzer:11}. Some works use the fact that both transforms can be interpreted as Fourier transforms over a suitable semigroup ring structure \cite{ABDALI1985257,LEHMANN197759}.

The first line of work focused on the powerset lattice $\myP = 2^{\A}$ of a set $\A$ ordered by inclusion. A fast zeta transform was discovered as early as ~\cite{yates1937design} (under the name factorial designs). In this case, both zeta and M\"obius transform can be done in $\On(|\myP|\log(|\myP|)$~\cite{kennes-moebius,moebius-computational-aspects}. If the result of the M\"obius transform of a powerset is $k$-sparse (i.e., only $k$ entries are nonzero), this can be reduced to $O(log(|\myP|\cdot |I(P)|)$, where $|I(P)|$ is the number order ideals of $\myP$ (i.e., the total number of antichains in $P$), or to $\On(\log(|\myP|) k^2)$ except for a set of Lebesgue measure zero.

For the larger, special class of lattices, \cite{few-irreducibles} develops an algorithm to compute both transforms in $\On(|\myP| \cdot \nu)$, where $\nu$ is the number of
\emph{irreducible} elements of $\myP$, by embedding the lattice into
a powerset of size $2^\nu$. Additionally, for a special family of lattices, they
achieve a complexity of $\On(|\Ered|))$. In~\cite{kaski2016fast} the latter
algorithm is extended to any semimodular lattice, and they observe that this complexity can be achieved for certain posets that are not lattices, namely those that can be equipped with an edge-rise labeling.
An edge-rising labeling is a function
$\lambda \colon \Ered \to \mathbb{Z}$ on the edges of the cover graph that
respects the order along each chain, i.e., $x_1 < \dots < x_n$ always implies
$\lambda((x_1,x_2)) < \dots < \lambda((x_{n-1},x_n))$. This class of posets is very restricted however; e.g., already the permutation lattice on three elements is not in it. The
lower bound $\Omega(|\Ered|)$ holds for general lattices~\cite{moebius-computational-aspects}, but not for posets as~\cite[Sec.~6]{few-irreducibles} shows using the construction
of~\cite{valiant1986negation}. For $k$-sparse results, $O(|\myP| k^3 + k^2)$ is possible~\cite{schapire1996learning}.

In the quest for fast Fourier transforms (FFTs) on various algebraic structures, specific cases of zeta transforms have been considered. The work in \cite{malandro2010fast} derives an FFT for the rook monoid
using, as one building block, a fast zeta transform on the poset of
all subgroups of the rook monoid ordered by subset inclusion.
In~\cite{malandro2015fourier} this approach is extended to general
finite inverse semigroups, now requiring a fast zeta transform on meet-semilattices, which are essentially equivalent to lattices in the case of finite $\myP$. Using general directed
multigraphs (called quivers) instead of posets enables FFTs
for any semisimple algebra~\cite{maslen2018efficient}.

In this paper we derive asymptotically and practically fast M\"obius and zeta transforms for {\em arbitrary} posets to enable the application to large-scale data.


\mypar{Applications} Posets, and the closely related directed acyclic graphs, are a natural model in many application scenarios. The classical application of the M\"obius inversion formula is to find solutions to problems in combinatorics~\cite{mob_appl_combinatorics, ordered-sets}, game theory~\cite{game_theory}, and in probability theory. In the latter, it can be used in the context of the Dempster-Shafer theory of
evidence to calculate belief values~\cite{dempster-shafer1,
  kennes-moebius,moebius-computational-aspects, dempster-shafer0} or to compute properties of Bayesian networks~\cite{koivisto-bayes-net}. In the context of probabilistic databases, it can efficiently evaluate
certain queries~\cite{p_databases_paper, p_database}. In control theory it has been used to design certain classes of optimal controllers~\cite{poset-decentralized-control,moebius-controller}.

Applications of the M\"obius inversion formula also extend to the
natural sciences. As an example, the M\"obius formula computes various
properties of a molecule by considering the poset induced by its atoms
and their bonds~\cite{cluster_expansion,molecule}.

Finally, \cite{lattice,discrete_meet_join_lattices} develops a linear signal processing framework,  for lattices, extended to arbitrary posets in \cite{seifert2022learning, Seifert22}, including prototypical applications, in which the M\"obius transform becomes the associated Fourier transform. The zeta transform plays the role of the inverse Fourier transform, where its columns yield the base frequencies.

\section{Background}\label{sec:back}

We provide the necessary background on partially ordered sets, their associated zeta and M\"obius transforms,
and their close relation to directed acyclic graphs. Then we explain the chain decomposition of DAGs, which is the primary tool that we use to construct fast zeta and M\"obius transforms. 

\mypar{Notation} Throughout this document, sets are denoted by calligraphic letters, and the elements of these sets are represented by lowercase letters. We use uppercase letters for matrices, and we represent the entry at row $i$ and column $j$ of a matrix $A$ by $a_{ij}$. 

\subsection{Posets and Directed Acyclic Graphs}

We provide background on posets and DAGs. For a more detailed introduction we refer to standard books such as \cite{stanley_2011}.

\mypar{Posets} A \emph{partially ordered set} (poset) is a set $\myP$ with a partial order $\partleq$. In this paper we assume $\myP$ is finite of size $\n$. Formally, $\partleq$ is a binary relation such that for all $x, y, z \in \myP$ the following holds:
\begin{enumerate}
    \item $x \partleq x$ (reflexivity),
    \item $x \partleq y$ and $y \partleq x$ implies $x = y$ (antisymmetry), 
    \item $x \partleq y$ and $y \partleq z$ implies $x \partleq z$
    (transitivity). 
\end{enumerate}
We write $x \partl y$ if $x \partleq y$ and $x \neq y$. We say that $y$ \emph{covers} $x$ if $x \partl y$ and there is no element in-between.

Posets can be topologically sorted into a list such that $x \partl y$ implies that $x$ comes after $y$ in that list. In the following we assume a fixed topologically sort and number the poset elements accordingly such that $\myP = \{x_1,\dots, x_{\n}\}$, i.e., $x_1$ is a largest, and $x_n$ a smallest elements in $\myP$. We choose a descending order since it more conveniently interfaces with the tools from \cite{klaus-simon} that we use later.

\mypar{DAGs and posets} A directed acyclic graph (DAG) is a directed graph $\G=(\myP,\E)$ that contains no cycles.

DAGs and posets are closely related: every DAG uniquely defines a partial order on its vertex set $\myP$, and, vice-versa, every poset can be represented by a DAG. More specifically, a DAG $\G=(\myP,\E)$ defines a partial order on $\myP$ by $x\partleq y$ if and only if there is a path from $y$ to $x$ in the DAG, i.e., $x$ is a successor of $y$. So the DAG traverses elements in descending order, a convention that is more convenient later than an ascending order, which can then be considered by simply flipping all edges.

It follows that all DAGs with the same transitive closure $(\myP, \Eclose)$ or the same transitive reduction $(\myP, \Ered)$ define the same partial order on $\myP$. The DAG with the minimal number of edges among those is $(\myP, \Ered)$, i.e., no edge is a transitive consequence of others.

Conversely, for a given a poset $\myP$, we can construct a DAG $\G=(\myP,\Ered)$ with $\Ered = \{ (x,y) \; | \; x \text{ covers } y\}$. This DAG is known as \emph{cover graph} of $\myP$ and it is in transitively reduced form. Another DAG representation of $\myP$ is the reachability graph $\G=(\myP,\Eclose)$ with $\Eclose = \{ (x,y) \; | \; y \partl x\}$. The cover graph is the transitive reduction of the reachability graph, and the reachability graph is the transitive closure of the cover graph. 

\begin{figure}
\centering
\hfill
    \begin{subfigure}[b]{0.32\textwidth}
   	\centering
   	\raisebox{0.05ex}{\includegraphics[scale=1.3]{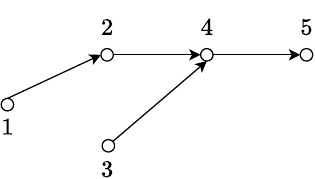}}
   	\caption{Transitive reduction (cover graph).}
\end{subfigure}
\hfill    
\begin{subfigure}[b]{0.32\textwidth}
    \centering
    \raisebox{2.52ex}{\includegraphics[scale=1.3]{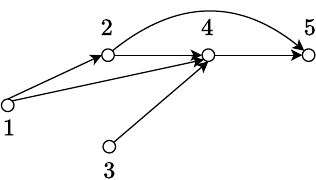}}
    \caption{DAG.}
\end{subfigure}
\hfill    
\begin{subfigure}[b]{0.32\textwidth}
	\centering
	\includegraphics[scale=1.3]{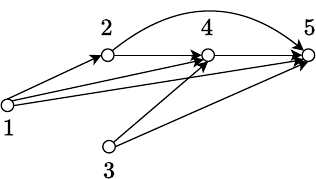}
	\caption{Transitive closure (reachability graph).}
\end{subfigure}
\hfill    
\caption{Example DAG~(b), its transitive reduction~(a), and its transitive closure~(c). All three DAGs define the same poset $\myP$.}\label{fig:dagclosered}
\end{figure}

An example is given in~\cref{fig:dagclosered}. All shown DAGs define the same poset $\myP = \{x_1,\dots,x_5\}$; the minimal one (in number of edges) is its cover graph and the maximal one its transitive closure. In this $\myP$, $x_1$ is the unique largest and $x_5$ the unique smallest element.

\mypar{Poset functions} The M\"obius inversion formula in~\cref{eq:inv-formula} operates on poset functions of the form $f:\myP\rightarrow\R$. Throughout this paper, we represent these maps using column vectors $\coord{f} \in \R^{\n}$ indexed by the elements of $\myP$, i.e., the $i$-th coordinate of $\coord{f}$ is $f(x_i)$. The set of these functions is an $n$-dimensional vector space.

\subsection{Zeta and M\"obius transform} 

We now describe in greater detail the inversion formula~\eqref{eq:inv-formula} already mentioned in~\cref{sec:intro}.

\begin{theorem}[M\"obius inversion formula~\cite{rota}]
\label{th:inv}
Given a poset $\myP$, and two poset functions $f, g: \myP \rightarrow \R$, we have
\begin{displaymath}
	g(x) = \sum_{y \partleq x} f(y)\quad\text{ if and only if }\quad f(x) = \sum_{y \partleq x} \mu(y,x)g(y),
\end{displaymath}
for all $x,y \in \myP$, and where $\mu : \myP \times \myP \rightarrow \Z$ is defined as~\cite{stanley_2011}
\begin{align*}\label{eq:MoebiusFunction}      
        \mu(x,x) &= 1, & \text{for } x \in \myP, \\
        \mu(x,y) &= - \sum_{x \partleq z \partl y} \mu(x,z), &
                     \text{for } x \not= y. 
\end{align*}
Consequently, $\mu(x,y) = 0$ for $x\npartleq y$.
\end{theorem}

Using the introduced vector notation, we can write Theorem~\ref{th:inv} equivalently as

\begin{corollary}
\label{co:inv-vec}
Let $Z = [\zeta(x,y)]_{x,y}\in\{0,1\}^{\n\times\n}$, where $\zeta(x,y) = 1$ if $y \partleq x$, and $0$ otherwise. Let $M = [\mu(y,x)]_{x,y}\in\Z^{\n\times\n}$. Then, for $\coord{f},\coord{g}\in\R^{\n}$,
\begin{equation}
\label{eq:inv-vec}
	\coord{g} = Z\coord{f}\quad\text{ if and only if }\quad \coord{f} = M\coord{g},
\end{equation}
which implies $Z^{-1} = M$.
\end{corollary}

If the poset is given by any DAG $\G = (\myP, \E)$, then $Z$ is the adjacency matrix of its transitive closure $(\myP, \Eclose)$. Note that due to the topological sort (from large to small), both $Z$ and $M$ are upper triangular. The two linear mappings in the theorem are called zeta and M\"obius transforms, respectively. Our goal is their efficient computation, given $\G$. 

A straightforward computation of the zeta transform would require a one-time computation of $\G$'s transitive closure. Then, assuming a sparse matrix storage format, the computation would require $O(\Eclose)$ space and time. For the M\"obius transform,  explicitly constructing $M$ is computationally intensive due to the recursive definition of $\mu(x,y)$~\cite{welsh2010matroid}. 
Thus, it is more efficient to compute $\coord{f} = M\coord{g}$ by solving the triangular linear system $Z\coord{f} = \coord{g}$, which again requires $O(\Eclose)$ space and time.

$Z$ only contains 0s and 1s and thus the zeta transform has significant redundancy. Specifically, every $2\times 2$ all-one submatrix represents a common subexpression that could be eliminated. The challenge is how to do so systematically, for any given poset. Our solution is based on the chain decomposition of DAGs that we discuss next, after a small example of the concepts introduced up to now.

\mypar{An example} We consider the DAG $\G = (\myP, \Ered)$ in~\cref{fig:dec-example}, which is in reduced form, i.e., the cover graph of a poset with twelve elements. Its adjacency matrix is given as (dots represent zeros)
\begin{displaymath}
A \in \{0,1\}^{12 \times 12}=\begin{psmallmatrix}
\phantom{1}\\
\zeroel&\zeroel&1&1&\zeroel&1&\zeroel&\zeroel&\zeroel&\zeroel&\zeroel&\zeroel\\
\zeroel&\zeroel&\zeroel&1&1&\zeroel&\zeroel&\zeroel&\zeroel&\zeroel&\zeroel&\zeroel\\
\zeroel&\zeroel&\zeroel&\zeroel&\zeroel&\zeroel&1&1&\zeroel&\zeroel&\zeroel&\zeroel\\
\zeroel&\zeroel&\zeroel&\zeroel&\zeroel&\zeroel&1&\zeroel&1&\zeroel&\zeroel&\zeroel\\
\zeroel&\zeroel&\zeroel&\zeroel&\zeroel&\zeroel&\zeroel&\zeroel&\zeroel&1&\zeroel&\zeroel\\
\zeroel&\zeroel&\zeroel&\zeroel&\zeroel&\zeroel&\zeroel&1&1&\zeroel&\zeroel&\zeroel\\
\zeroel&\zeroel&\zeroel&\zeroel&\zeroel&\zeroel&\zeroel&\zeroel&\zeroel&1&\zeroel&\zeroel\\
\zeroel&\zeroel&\zeroel&\zeroel&\zeroel&\zeroel&\zeroel&\zeroel&\zeroel&\zeroel&1&\zeroel\\
\zeroel&\zeroel&\zeroel&\zeroel&\zeroel&\zeroel&\zeroel&\zeroel&\zeroel&\zeroel&\zeroel&1\\
\zeroel&\zeroel&\zeroel&\zeroel&\zeroel&\zeroel&\zeroel&\zeroel&\zeroel&\zeroel&1&1\\
\zeroel&\zeroel&\zeroel&\zeroel&\zeroel&\zeroel&\zeroel&\zeroel&\zeroel&\zeroel&\zeroel&\zeroel\\
\zeroel&\zeroel&\zeroel&\zeroel&\zeroel&\zeroel&\zeroel&\zeroel&\zeroel&\zeroel&\zeroel&\zeroel\\
\phantom{1}
\end{psmallmatrix}.
\end{displaymath}
The associated $Z$ and $M$ matrices are 
\begin{equation}\label{eq:zm}
Z=\begin{psmallmatrix}
\phantom{1}\\
1&\zeroel&1&1&\zeroel&1&1&1&1&1&1&1\\
\zeroel&1&\zeroel&1&1&\zeroel&1&\zeroel&1&1&1&1\\
\zeroel&\zeroel&1&\zeroel&\zeroel&\zeroel&1&1&\zeroel&1&1&1\\
\zeroel&\zeroel&\zeroel&1&\zeroel&\zeroel&1&\zeroel&1&1&1&1\\
\zeroel&\zeroel&\zeroel&\zeroel&1&\zeroel&\zeroel&\zeroel&\zeroel&1&1&1\\
\zeroel&\zeroel&\zeroel&\zeroel&\zeroel&1&\zeroel&1&1&\zeroel&1&1\\
\zeroel&\zeroel&\zeroel&\zeroel&\zeroel&\zeroel&1&\zeroel&\zeroel&1&1&1\\
\zeroel&\zeroel&\zeroel&\zeroel&\zeroel&\zeroel&\zeroel&1&\zeroel&\zeroel&1&\zeroel\\
\zeroel&\zeroel&\zeroel&\zeroel&\zeroel&\zeroel&\zeroel&\zeroel&1&\zeroel&\zeroel&1\\
\zeroel&\zeroel&\zeroel&\zeroel&\zeroel&\zeroel&\zeroel&\zeroel&\zeroel&1&1&1\\
\zeroel&\zeroel&\zeroel&\zeroel&\zeroel&\zeroel&\zeroel&\zeroel&\zeroel&\zeroel&1&\zeroel\\
\zeroel&\zeroel&\zeroel&\zeroel&\zeroel&\zeroel&\zeroel&\zeroel&\zeroel&\zeroel&\zeroel&1\\
\phantom{1}
\end{psmallmatrix}, \
M=\begin{psmallmatrix}
\phantom{1}\\
\phantom{-}1&&-1&-1&&-1&\phantom{-}1&\phantom{-}1&\phantom{-}1&\zeroelph&\zeroelph&\zeroelph\\
\zeroelph&\phantom{-}1&\zeroelph&-1&-1&\zeroelph&\zeroelph&\zeroelph&\zeroelph&\phantom{-}1&\zeroelph&\zeroelph\\
\zeroelph&\zeroelph&\phantom{-}1&\zeroelph&\zeroelph&\zeroelph&-1&-1&\zeroelph&\zeroelph&\phantom{-}1&\zeroelph\\
\zeroelph&\zeroelph&\zeroelph&\phantom{-}1&\zeroelph&\zeroelph&-1&\zeroelph&-1&\zeroelph&\zeroelph&\phantom{-}1\\
\zeroelph&\zeroelph&\zeroelph&\zeroelph&\phantom{-}1&\zeroelph&\zeroelph&\zeroelph&\zeroelph&-1&\zeroelph&\zeroelph\\
\zeroelph&\zeroelph&\zeroelph&\zeroelph&\zeroelph&\phantom{-}1&\zeroelph&-1&-1&\zeroelph&\zeroelph&\zeroelph\\
\zeroelph&\zeroelph&\zeroelph&\zeroelph&\zeroelph&\zeroelph&\phantom{-}1&\zeroelph&\zeroelph&-1&\zeroelph&\zeroelph\\
\zeroelph&\zeroelph&\zeroelph&\zeroelph&\zeroelph&\zeroelph&\zeroelph&\phantom{-}1&\zeroelph&\zeroelph&-1&\zeroelph\\
\zeroelph&\zeroelph&\zeroelph&\zeroelph&\zeroelph&\zeroelph&\zeroelph&\zeroelph&\phantom{-}1&\zeroelph&\zeroelph&-1\\
\zeroelph&\zeroelph&\zeroelph&\zeroelph&\zeroelph&\zeroelph&\zeroelph&\zeroelph&\zeroelph&\phantom{-}1&-1&-1\\
\zeroelph&\zeroelph&\zeroelph&\zeroelph&\zeroelph&\zeroelph&\zeroelph&\zeroelph&\zeroelph&\zeroelph&\phantom{-}1&\zeroelph\\
\zeroelph&\zeroelph&\zeroelph&\zeroelph&\zeroelph&\zeroelph&\zeroelph&\zeroelph&\zeroelph&\zeroelph&\zeroelph&\phantom{-}1\\
\phantom{1}
\end{psmallmatrix},
\end{equation}
where $Z$ is the adjacency matrix of $\G = (\myP, \Eclose)$. Note that although $A$ is sparse in this example, the number of nonzero elements in $Z$ and $M$ increases significantly. For example
$$
g(x_1) = \sum_{x_i\partleq x_1}f(x_i) = f(x_1) + f(x_3) + f(x_4) + f(x_6) + \dots + f(x_{12}).
$$
Note that the entire computation $\coord{g}=Z\coord{f}$ requires, for example, the repeated addition $f(x_{11}) + f(x_{12})$, which thus is a common subexpression.

\subsection{Chain Decomposition}
\label{subsec:cd} 

Our approach first requires a chain decomposition of the given DAG:

\begin{definition}[Chain decomposition]\label{def:cd}
    Let $G=(\myP,\E)$ be a DAG. A \emph{chain decomposition} $\Zc = \{ \Zc_1, \dots, \Zc_k \}$ of $\G$ is a partition of $\myP$ such that every $\Zc_i$ is a totally ordered subset of $\myP$. 
\end{definition}

\begin{figure}
    \centering
    \hfill    
    \begin{subfigure}[b]{0.45\textwidth}
        \centering
        \raisebox{0.98ex}{\includegraphics[scale=1.3]{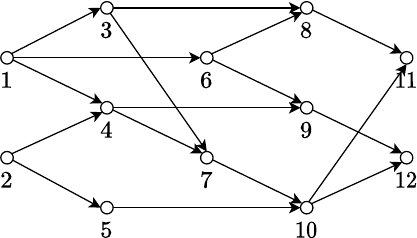}}
        \caption{A poset as cover graph}
    \end{subfigure}
    \hfill
    \begin{subfigure}[b]{0.45\textwidth}
        \centering
        \includegraphics[scale=1.3]{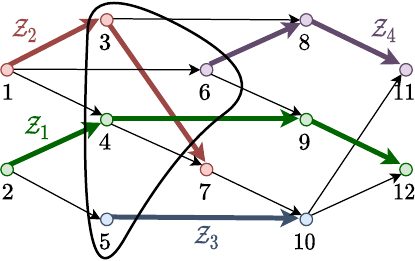}
        \caption{An associated chain decomposition}
    \end{subfigure}
    \hfill    
    \caption{Example of a poset represented by its cover graph~(a) and one chain decomposition $\{\Zc_1, \Zc_2, \Zc_3, \Zc_4\}$
      in~(b). The chains are highlighted in bold and different colors. A longest antichain, $\{3, 4, 5, 6\}$, is circled in black.}\label{fig:dec-example}
\end{figure}


\cref{fig:dec-example} shows an example of a chain decomposition.

\mypar{Dilworth's theorem} We will use the following result from~\cite{dilworth}, which shows that the minimal number of chains in a decomposition is given by $\width(\myP)$, which is the size of the longest \emph{antichain} in $\myP$. An antichain is a set of pairwise non-comparable elements. In our example in~\cref{fig:dec-example}, a longest antichain is the circled set of vertices $\{3,4,5,6\}$.

\begin{theorem}[Dilworth's Chain Decomposition
    Theorem~\cite{dilworth}]\label{th:dilworth} 
    A minimal chain decomposition of a DAG $\G = (\myP, \E)$ has size $k = \width(\myP)$.
\end{theorem}

In the following we will always work with minimal chain decompositions of the DAG, which can be computed in $\On(\n \cdot |\E|)$ by solving a maximum matching problem \cite{fulkerson}. In our experiments we also use the algorithm from \cite{max-flow}, which is often faster in practice despite a worst case runtime of $\On(n^3)$.

\subsection{Computing reachability} 
\label{subsec:reach}
Assuming a chain decomposition, we define the function $\next(x)$ that returns the vertex following $x$ on its chain. For the last element in a chain, we set $\next(x) = x$.

Using chain decompositions, \cite{klaus-simon} presents a method to efficiently answer queries of the form ``$x\partleq y$?'' by computing the reachability set $\closure{x}= \{ y \in \myP \; | \; y \partleq x\}$. This is done by defining three maps associated with a chain decomposition $\Zc = \{\Zc_1, \dots, \Zc_k\}$:
\begin{itemize}
    \item $\id$ maps each vertex in $\myP$ to its chain identifier:
    \begin{align*}  
      \id: \myP &\rightarrow \{1,\dots,k\}, \\
      x &\mapsto i \text{ where } x \in \Zc_i,
    \end{align*}
    \item $\niv_j$ maps each vertex to the smallest element in $\closure{x} \cap \Zc_j$:
    \begin{align*}  
      \niv_j: \myP &\rightarrow \myP \cup \{\infty\}, \\
      v &\mapsto 
      	\begin{cases}
			\max(\closure{x} \cap \Zc_j), \ &\closure{x} \cap \Zc_j \neq \emptyset,\\
			\infty, \ &\text{otherwise},
		\end{cases}
    \end{align*}
    \item $\niv$ collects the values from all $\niv_j$:
    i.e., 
    \begin{align*}  
      \niv: \myP &\rightarrow 2^{\myP \cup \{\infty\}}, \\
      x &\mapsto \{\niv_j(x) | 1 \leq j \leq k\}.
    \end{align*}
    Note that by construction $v \in \niv(v)$.
\end{itemize}

In this paper, we utilize these three maps in combination with the chain decomposition to efficiently compute both zeta and M\"obius transform. Note that we store at most $k$ $\niv_j$ values for each element since the DAG is decomposed in $k$ chains~\cref{th:dilworth}. Hence, our approach requires $\On(nk)$ memory.

We can think of $\niv$ for a DAG as a matrix $N \in \myP^{\n \times k}$ where each entry is $n_{ij} = \niv_j(i)$. In practice, we consider $N$ as a sparse matrix by not storing the $\infty$ elements, and we refer to the number of non-infinite entries with $\nivsize$. For the DAG in~\cref{fig:dec-example}, $N\in\myP^{12\times 4}$ and $\nivsize = 32$:

\def\tmp{%
  \begin{xsmallmatrix}{1em}
\?\phantom{4}\?&\?\phantom{4}\?&\?\phantom{40}\?&\?\phantom{4}\?\\
4&1&10&6\\
2&7&5&11\\
12&3&10&8\\
4&7&10&11\\
12&\infty&5&11\\
9&\infty&\infty&6\\
12&7&10&11\\
\infty&\infty&\infty&8\\
9&\infty&\infty&\infty\\
12&\infty&10&11\\
\infty&\infty&\infty&11\\
12&\infty&\infty&\infty\\
\phantom{1}
 \end{xsmallmatrix}
}%

\begin{displaymath}
N=
\stackMath\def\stackalignment{r}%
{\def\?{\kern2pt}
  \stackon%
    {\!\left(\tmp\right)}%
    {\stackon[1pt]{\phantom{\smash{\tmp\mkern -30mu}}}{%
    \begin{smallmatrix}\?\Zc_1\?&\? \, \Zc_2\?\?&\? \, \Zc_3\?\?&\? \, \Zc_4\end{smallmatrix}}\mkern 15mu}}%
    .
\end{displaymath}

\begin{figure}
    \centering
    \includegraphics[scale=1.3]{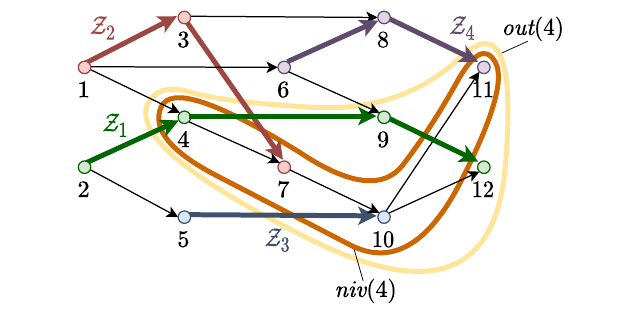}
    \caption{Example of $\niv$ (circled in red) and $\textit{out}$ (circled in yellow) for vertex~4.}
    \label{fig:niv-ex}
\end{figure}

For example, if we consider vertex $4$ in~\cref{fig:niv-ex}, we have
\begin{displaymath}
	\niv_{1}(4) = 4, \ \niv_{2}(4)=7, \ \niv_{3}(4)=10, \ \niv_{4}(4) = 11.
\end{displaymath}

\mypar{Example} Given $\id$ and $\niv$, the reachability set $\closure{x}$ of a vertex $x$ can be computed. We consider first an example: $\closure{4}$ in the DAG in~\cref{fig:dec-example}, assuming $N$ above has been computed.

We have $\niv(4) = \{ 4, 7, 10 , 11\}$ and compute $\closure{4}$ by traversing each chain $j$ (using the function $\next$) starting from $\niv_j(4)$. In this case, this yields $\{4, 9, 12\}$ by traversing $\Zc_1$, $\{ 7\}$ by traversing $\Zc_2$, $\{ 10\}$ by traversing $\Zc_3$, and $\{ 11 \}$ by traversing $\Zc_4$. The (necessarily) disjoint union of these sets yields $\closure{4} = \{ 4, 7, 9, 10, 11, 12\}$, as visualized in~\cref{fig:niv-ex}.

\mypar{General case} \cite{klaus-simon} generalizes the process of the example with the following theorem. For a proof, see~\cite{klaus-simon}.

\begin{theorem}\label{thm:reach}
Given a DAG $\G=(\myP,\E)$ and a chain decomposition 
$\Zc = \{ \Zc_1,$ $\dots,$ $\Zc_k\}$, for every $x$ in $\myP$
\begin{displaymath}
    \closure{x} = \mathop{\dot{\bigcup}}\limits_{1 \leq j \leq k} \{z \in \Zc_j | z \partleq \niv_j(x) \},
\end{displaymath}
where the dot denotes a disjoint union.
\end{theorem}

As mentioned, $\id$ and $\niv$ require $\On(\n k)$ memory and the computation of the union requires no additional memory. Further, computing $\closure{x}$ from $\niv$ can be done in $\On(\n)$ time as we visit each vertex at most once. A standard BFS (breadth-first search) approach to compute $\closure{x}$ would require $\On(\n + |\E|)$ operations for each vertex.

It remains the efficient construction of the $\niv$-matrix $N$. This is done in $\On(|\E| + |\Ered| \cdot k)$ using the following theorem from~\cite{klaus-simon}.

\begin{theorem}\label{thm:computenivj}
Given a DAG $\G=(\myP,\E)$ and a chain decomposition $\Zc = \{ \Zc_1,$ $\dots,$ $\Zc_k\}$, for every $x$ in $\myP$ and for every $j$ such that $1 \leq j \leq k$
\begin{displaymath}
    \niv_j(x) = 
    	\begin{cases}
    		\max(\{\niv_j(z)|z \in \adj(x)\}) & j\neq \id(x),\\
    		x & \text{otherwise}.
		\end{cases}
\end{displaymath}
\end{theorem}

Here, $\adj(x)$ is the set of successors of $x$, i.e., the set of elements covered by $x$.


\section{Main results}
\label{sec:main}

We describe our approach to compute the zeta and M\"obius transforms on poset functions, which we represent as vectors indexed by the elements of the poset. Existing methods operate on particular instances of posets called lattices~\cite{few-irreducibles,kennes-moebius,moebius-computational-aspects}, which are posets where all pairs of elements have both a unique lower and unique upper bound. Our approach is not limited to specific poset structures and can operate on any DAGs.

We recall that a naive implementation of these transformations requires computing and storing the $Z$ and $M$ transformation matrices of~\eqref{eq:inv-vec}. The main limiting factor of this approach is its prohibitively expensive $\On(\n^2)$ memory requirement for both matrices. Our approach utilizes the ($\On(nk)$ memory) chain decomposition instead, as described in~\cref{subsec:cd}.

We show that our strategy factorizes the $Z$ matrix into two distinct sparse matrices, which reduces the memory and time complexity of the transformations from $\On(\n^2)$ of the naive implementation, to $\On(nk)$, with $k$ being the width of the input DAG. From our experiments, $\On(nk)$ is a loose bound, and our implementation is often faster in practice.

\mypar{Fast zeta transform}
Recall from~\cref{co:inv-vec} that the zeta transform of a vector $\coord{x}$ indexed by a poset $\myP$ is the matrix-vector multiplication $\coord{y} = Z \coord{x}$, where $Z$ is the adjacency matrix of the transitive closure of $\myP$. More specifically, each coordinate of $\coord{y}$ is computed as
\begin{equation} \label{eq:zeta}
    y_i = \sum_{z \partleq x_i}z,
\end{equation}
by summing all the elements $z$ such that $z \partleq x_i$ in $\myP$. 

We use the chain decomposition to rewrite~\eqref{eq:zeta} as
\begin{equation} \label{eq:zeta-mid}
    y_i = \sum_{1 \leq q \leq k}\bigg{(}\sum_{\substack{z \partleq \niv_q(x_i)\\ q = \id(z)}} z\bigg{)}.
\end{equation}
The only difference between~\eqref{eq:zeta} and~\eqref{eq:zeta-mid} is that we are now explicitly adding elements by traversing chains. This solution requires only the chain decomposition, $\id$, and $\niv$. Hence, the total memory complexity is reduced from $\On(\n^2)$ to $\On(\n k)$, while maintaining the $\On(\n^2)$ bound on the time complexity. To also reduce the number of operations we need one final observation: since the chains are disjoint by construction, the innermost sum of~\eqref{eq:zeta-mid} iterates over the same elements multiple times and can be substituted by a linear map. In particular, we represent this linear map as a vector $\coord{c}$, indexed by the poset $\myP$, where each $i$-th component is defined by
\begin{equation}\label{eq:zeta_cache}
    c_i = \sum_{\substack{z \partleq x_i \\ \id(z) = \id(x_i)}} z.
\end{equation}
Alternatively, we can define $\coord{c}$ recursively by using the $\next$ map as
\begin{displaymath}
c_i = \begin{cases}
x_i + c_{\next(x_i)}, & x_i \neq \next(x_i),\\
x_i, & \text{otherwise.}
\end{cases}
\end{displaymath}
Finally, we simplify~\eqref{eq:zeta-mid} by summing the entries of $\coord{c}$ instead of iterating over the full transitive closure. We obtain the following equation
\begin{align}
    y_i &= \sum_{1 \leq q \leq k}\bigg{(}\sum_{\substack{j \partleq \niv_q(i)\\ q = \id(j)}} x_j\bigg{)} 
    = \sum_{1 \leq q \leq k} c_{\niv_q(i)} 
    = \sum_{j \in \niv(i)} c_j\label{eq:zeta-fast}.
\end{align}
From this formulation we notice that time complexity is indeed linear in $\nivsize = \On(\n k)$.

Using matrix notation, the fast zeta transform in~\eqref{eq:zeta-fast} corresponds to a factorization
\begin{equation}\label{eq:zfac}
Z = UV,
\end{equation}
with sparse $U$ and $V$. In particular, the matrix $U \in \{0,1\}^{\n \times \n}$ is such that each element $u_{ij} = 1$ if $x_j \in \niv(x_i)$ and $0$ otherwise, and the matrix $V \in \{0,1\}^{\n \times \n}$ is the adjacency matrix of the transitive closure of the chain decomposition. Furthermore, the matrix $V$ is a block-diagonal matrix $B$, conjugated by a permutation, in which each block is the zeta transform of a chain of the DAG. Hence, the factorization can also be written as
\begin{equation}\label{eq:finz}
	Z = UP^{-1}BP.
\end{equation}
where $P$ is a permutation matrix.

Using this factorization, we analyze the number of operations of this fast zeta transform. The first transformation, $\coord{c} = V \coord{x}$, requires $\n - k$ additions, as we are visiting each element of $\myP$ only once while traversing the $k$ chains. The second transformation, $\coord{y} = U \coord{c}$, requires $\nivsize - \n$ additions. 
Thus, the total number of operations is $\n - k + \nivsize - \n = \nivsize - k \leq \n k - k$.

\mypar{Example} We continue the example from~\cref{fig:dec-example,fig:niv-ex}. The factorization $Z = UV$ for $Z$ in \eqref{eq:zm} becomes
\begin{equation}\label{eq:zuv}
Z=\begin{psmallmatrix}
\phantom{1}\\
1&\zeroel&1&1&\zeroel&1&1&1&1&1&1&1\\
\zeroel&1&\zeroel&1&1&\zeroel&1&\zeroel&1&1&1&1\\
\zeroel&\zeroel&1&\zeroel&\zeroel&\zeroel&1&1&\zeroel&1&1&1\\
\zeroel&\zeroel&\zeroel&1&\zeroel&\zeroel&1&\zeroel&1&1&1&1\\
\zeroel&\zeroel&\zeroel&\zeroel&1&\zeroel&\zeroel&\zeroel&\zeroel&1&1&1\\
\zeroel&\zeroel&\zeroel&\zeroel&\zeroel&1&\zeroel&1&1&\zeroel&1&1\\
\zeroel&\zeroel&\zeroel&\zeroel&\zeroel&\zeroel&1&\zeroel&\zeroel&1&1&1\\
\zeroel&\zeroel&\zeroel&\zeroel&\zeroel&\zeroel&\zeroel&1&\zeroel&\zeroel&1&\zeroel\\
\zeroel&\zeroel&\zeroel&\zeroel&\zeroel&\zeroel&\zeroel&\zeroel&1&\zeroel&\zeroel&1\\
\zeroel&\zeroel&\zeroel&\zeroel&\zeroel&\zeroel&\zeroel&\zeroel&\zeroel&1&1&1\\
\zeroel&\zeroel&\zeroel&\zeroel&\zeroel&\zeroel&\zeroel&\zeroel&\zeroel&\zeroel&1&\zeroel\\
\zeroel&\zeroel&\zeroel&\zeroel&\zeroel&\zeroel&\zeroel&\zeroel&\zeroel&\zeroel&\zeroel&1\\
\phantom{1}
\end{psmallmatrix}
=
\begin{psmallmatrix}
\phantom{1}\\
1&\zeroel&\zeroel&1&\zeroel&1&\zeroel&\zeroel&\zeroel&1&\zeroel&\zeroel\\
\zeroel&1&\zeroel&\zeroel&1&\zeroel&1&\zeroel&\zeroel&\zeroel&1&\zeroel\\
\zeroel&\zeroel&1&\zeroel&\zeroel&\zeroel&\zeroel&1&\zeroel&1&\zeroel&1\\
\zeroel&\zeroel&\zeroel&1&\zeroel&\zeroel&1&\zeroel&\zeroel&1&1&\zeroel\\
\zeroel&\zeroel&\zeroel&\zeroel&1&\zeroel&\zeroel&\zeroel&\zeroel&\zeroel&1&1\\
\zeroel&\zeroel&\zeroel&\zeroel&\zeroel&1&\zeroel&\zeroel&1&\zeroel&\zeroel&\zeroel\\
\zeroel&\zeroel&\zeroel&\zeroel&\zeroel&\zeroel&1&\zeroel&\zeroel&1&1&1\\
\zeroel&\zeroel&\zeroel&\zeroel&\zeroel&\zeroel&\zeroel&1&\zeroel&\zeroel&\zeroel&\zeroel\\
\zeroel&\zeroel&\zeroel&\zeroel&\zeroel&\zeroel&\zeroel&\zeroel&1&\zeroel&\zeroel&\zeroel\\
\zeroel&\zeroel&\zeroel&\zeroel&\zeroel&\zeroel&\zeroel&\zeroel&\zeroel&1&1&1\\
\zeroel&\zeroel&\zeroel&\zeroel&\zeroel&\zeroel&\zeroel&\zeroel&\zeroel&\zeroel&1&\zeroel\\
\zeroel&\zeroel&\zeroel&\zeroel&\zeroel&\zeroel&\zeroel&\zeroel&\zeroel&\zeroel&\zeroel&1\\
\phantom{1}
\end{psmallmatrix}
\cdot
\begin{psmallmatrix}
\phantom{1}\\
1&\zeroel&1&\zeroel&\zeroel&\zeroel&1&\zeroel&\zeroel&\zeroel&\zeroel&\zeroel\\
\zeroel&1&\zeroel&1&\zeroel&\zeroel&\zeroel&\zeroel&1&\zeroel&\zeroel&1\\
\zeroel&\zeroel&1&\zeroel&\zeroel&\zeroel&1&\zeroel&\zeroel&\zeroel&\zeroel&\zeroel\\
\zeroel&\zeroel&\zeroel&1&\zeroel&\zeroel&\zeroel&\zeroel&1&\zeroel&\zeroel&1\\
\zeroel&\zeroel&\zeroel&\zeroel&1&\zeroel&\zeroel&\zeroel&\zeroel&1&\zeroel&\zeroel\\
\zeroel&\zeroel&\zeroel&\zeroel&\zeroel&1&\zeroel&1&\zeroel&\zeroel&1&\zeroel\\
\zeroel&\zeroel&\zeroel&\zeroel&\zeroel&\zeroel&1&\zeroel&\zeroel&\zeroel&\zeroel&\zeroel\\
\zeroel&\zeroel&\zeroel&\zeroel&\zeroel&\zeroel&\zeroel&1&\zeroel&\zeroel&1&\zeroel\\
\zeroel&\zeroel&\zeroel&\zeroel&\zeroel&\zeroel&\zeroel&\zeroel&1&\zeroel&\zeroel&1\\
\zeroel&\zeroel&\zeroel&\zeroel&\zeroel&\zeroel&\zeroel&\zeroel&\zeroel&1&\zeroel&\zeroel\\
\zeroel&\zeroel&\zeroel&\zeroel&\zeroel&\zeroel&\zeroel&\zeroel&\zeroel&\zeroel&1&\zeroel\\
\zeroel&\zeroel&\zeroel&\zeroel&\zeroel&\zeroel&\zeroel&\zeroel&\zeroel&\zeroel&\zeroel&1\\
\phantom{1}
\end{psmallmatrix}.
\end{equation}
The interpretation of these two matrices is fairly straightforward. For example, the nonzero elements in the first row of $U$, correspond to the values in the first row of the $N$ matrix, which is $(4, 1, 10, 6)$. Similarly, the nonzero elements in the first row of $V$ correspond to the transitive closure of $\Zc_2 = \{1, 3, 7\}$. Note that we can obtain the entries of $V$ by using the $\next$ function defined in~\cref{subsec:reach}. Indeed, for the first row of $V$ we have $\next(1) = 3$, $\next(3) = 7$ and $\next(7) = 7$. We use this relation in our algorithms.

Additionally, after reordering rows and columns with the same permutation $P$, $V$ can be written as a block matrix
\begin{equation}\label{eq:exb}
PVP^{-1} = B=\begin{psmallmatrix}
\phantom{1}\\
1&1&1&1&\zeroel&\zeroel&\zeroel&\zeroel&\zeroel&\zeroel&\zeroel&\zeroel\\
\zeroel&1&1&1&\zeroel&\zeroel&\zeroel&\zeroel&\zeroel&\zeroel&\zeroel&\zeroel\\
\zeroel&\zeroel&1&1&\zeroel&\zeroel&\zeroel&\zeroel&\zeroel&\zeroel&\zeroel&\zeroel\\
\zeroel&\zeroel&\zeroel&1&\zeroel&\zeroel&\zeroel&\zeroel&\zeroel&\zeroel&\zeroel&\zeroel\\
\zeroel&\zeroel&\zeroel&\zeroel&1&1&1&\zeroel&\zeroel&\zeroel&\zeroel&\zeroel\\
\zeroel&\zeroel&\zeroel&\zeroel&\zeroel&1&1&\zeroel&\zeroel&\zeroel&\zeroel&\zeroel\\
\zeroel&\zeroel&\zeroel&\zeroel&\zeroel&\zeroel&1&\zeroel&\zeroel&\zeroel&\zeroel&\zeroel\\
\zeroel&\zeroel&\zeroel&\zeroel&\zeroel&\zeroel&\zeroel&1&1&\zeroel&\zeroel&\zeroel\\
\zeroel&\zeroel&\zeroel&\zeroel&\zeroel&\zeroel&\zeroel&\zeroel&1&\zeroel&\zeroel&\zeroel\\
\zeroel&\zeroel&\zeroel&\zeroel&\zeroel&\zeroel&\zeroel&\zeroel&\zeroel&1&1&1\\
\zeroel&\zeroel&\zeroel&\zeroel&\zeroel&\zeroel&\zeroel&\zeroel&\zeroel&\zeroel&1&1\\
\zeroel&\zeroel&\zeroel&\zeroel&\zeroel&\zeroel&\zeroel&\zeroel&\zeroel&\zeroel&\zeroel&1\\
\phantom{1}
\end{psmallmatrix}.
\end{equation}
Each block corresponds to the zeta transform of a total order, that, in our case, is given by the four chains $\Zc_1, \Zc_2, \Zc_3, \Zc_4$. For this example, the permutation $P = (1, 2, 4, 12, 11, 8, 5)(3, 9, 10, 6)$, written in cycle notation. Note that $B$ is not unique, but depends on the order in which we place the blocks. Thus the final factorization is $Z = UP^{-1}BP$. Multiplying a vector by $P^{-1}BP$ requires 8 additions. Multiplying the result by $U$ and additional 20 for a total of 28 additions. Multiplying by $Z$ directly requires 39 additions.


\mypar{Fast M\"obius transform}
Our goal is to compute the M\"obius transform $M\coord{x} = \coord{y}$ without explicitly constructing the matrix $M = Z^{-1}$. We do this by solving the equivalent triangular linear system $\coord{x} = Z\coord{y}$. Using \eqref{eq:finz}, we do this in two steps. First, we solve the system $\coord{x} = U\coord{y'}$. Second, we compute the final result as $\coord{y} = P^{-1}B^{-1}P\coord{y'}$. For the latter we use that the inverse of a triangular matrix of ones is the identity matrix with additional $-1$'s on the upper diagonal. Therefore, each row of $P^{-1}B^{-1}P$ will amount to at most a single subtraction.

As before, since we store $Z$ and $P^{-1}B^{-1}P$ in sparse format and they are upper triangular, the number of operations is the number of nonzero elements in the matrices. In particular, we need $\nivsize - \n$ subtractions for the triangular linear solve $\coord{x} = U\coord{y'}$ via backsubstitution, and $\n - k$ subtractions for $\coord{y} = P^{-1}B^{-1}P\coord{y'}$, for a total of $\nivsize - \n + \n - k = \nivsize - k \leq \n k - k$ operations. Thus, both our fast zeta transform and fast M\"obius transform require $\nivsize - k$ operations.

\mypar{An example}
The matrix $U$ to solve the linear system is as in \eqref{eq:zuv}. Here, we show how to construct the transformation matrix $P^{-1}B^{-1}P$. As stated, it is easy to find the inverse of the block triangular matrix $B$, given by
\begin{displaymath}
B^{-1} =\begin{psmallmatrix}
\phantom{1}\\
\phantom{-}1&-1&\zeroelph&\zeroelph&\zeroelph&\zeroelph&\zeroelph&\zeroelph&\zeroelph&\zeroelph&\zeroelph&\zeroelph\\
\zeroelph&\phantom{-}1&-1&\zeroelph&\zeroelph&\zeroelph&\zeroelph&\zeroelph&\zeroelph&\zeroelph&\zeroelph&\zeroelph\\
\zeroelph&\zeroelph&\phantom{-}1&-1&\zeroelph&\zeroelph&\zeroelph&\zeroelph&\zeroelph&\zeroelph&\zeroelph&\zeroelph\\
\zeroelph&\zeroelph&\zeroelph&\phantom{-}1&\zeroelph&\zeroelph&\zeroelph&\zeroelph&\zeroelph&\zeroelph&\zeroelph&\zeroelph\\
\zeroelph&\zeroelph&\zeroelph&\zeroelph&\phantom{-}1&-1&\zeroelph&\zeroelph&\zeroelph&\zeroelph&\zeroelph&\zeroelph\\
\zeroelph&\zeroelph&\zeroelph&\zeroelph&\zeroelph&\phantom{-}1&-1&\zeroelph&\zeroelph&\zeroelph&\zeroelph&\zeroelph\\
\zeroelph&\zeroelph&\zeroelph&\zeroelph&\zeroelph&\zeroelph&\phantom{-}1&\zeroelph&\zeroelph&\zeroelph&\zeroelph&\zeroelph\\
\zeroelph&\zeroelph&\zeroelph&\zeroelph&\zeroelph&\zeroelph&\zeroelph&\phantom{-}1&-1&\zeroelph&\zeroelph&\zeroelph\\
\zeroelph&\zeroelph&\zeroelph&\zeroelph&\zeroelph&\zeroelph&\zeroelph&\zeroelph&\phantom{-}1&\zeroelph&\zeroelph&\zeroelph\\
\zeroelph&\zeroelph&\zeroelph&\zeroelph&\zeroelph&\zeroelph&\zeroelph&\zeroelph&\zeroelph&\phantom{-}1&-1&\zeroelph\\
\zeroelph&\zeroelph&\zeroelph&\zeroelph&\zeroelph&\zeroelph&\zeroelph&\zeroelph&\zeroelph&\zeroelph&\phantom{-}1&-1\\
\zeroelph&\zeroelph&\zeroelph&\zeroelph&\zeroelph&\zeroelph&\zeroelph&\zeroelph&\zeroelph&\zeroelph&\zeroelph&\phantom{-}1\\
\phantom{1}
\end{psmallmatrix},
\end{displaymath}
which has the same block structure as $B$ in \eqref{eq:exb}. This matrix is then permuted using the previous permutation matrix $P$ to obtain
\begin{displaymath}
V^{-1} = P^{-1}B^{-1}P =\begin{psmallmatrix}
\phantom{1}\\
\phantom{-}1&\zeroelph&-1&\zeroelph&\zeroelph&\zeroelph&\zeroelph&\zeroelph&\zeroelph&\zeroelph&\zeroelph&\zeroelph\\
\zeroelph&\phantom{-}1&\zeroelph&-1&\zeroelph&\zeroelph&\zeroelph&\zeroelph&\zeroelph&\zeroelph&\zeroelph&\zeroelph\\
\zeroelph&\zeroelph&\phantom{-}1&\zeroelph&\zeroelph&\zeroelph&-1&\zeroelph&\zeroelph&\zeroelph&\zeroelph&\zeroelph\\
\zeroelph&\zeroelph&\zeroelph&\phantom{-}1&\zeroelph&\zeroelph&\zeroelph&\zeroelph&-1&\zeroelph&\zeroelph&\zeroelph\\
\zeroelph&\zeroelph&\zeroelph&\zeroelph&\phantom{-}1&\zeroelph&\zeroelph&\zeroelph&\zeroelph&-1&\zeroelph&\zeroelph\\
\zeroelph&\zeroelph&\zeroelph&\zeroelph&\zeroelph&\phantom{-}1&\zeroelph&-1&\zeroelph&\zeroelph&\zeroelph&\zeroelph\\
\zeroelph&\zeroelph&\zeroelph&\zeroelph&\zeroelph&\zeroelph&\phantom{-}1&\zeroelph&\zeroelph&\zeroelph&\zeroelph&\zeroelph\\
\zeroelph&\zeroelph&\zeroelph&\zeroelph&\zeroelph&\zeroelph&\zeroelph&\phantom{-}1&\zeroelph&\zeroelph&-1&\zeroelph\\
\zeroelph&\zeroelph&\zeroelph&\zeroelph&\zeroelph&\zeroelph&\zeroelph&\zeroelph&\phantom{-}1&\zeroelph&\zeroelph&-1\\
\zeroelph&\zeroelph&\zeroelph&\zeroelph&\zeroelph&\zeroelph&\zeroelph&\zeroelph&\zeroelph&\phantom{-}1&\zeroelph&\zeroelph\\
\zeroelph&\zeroelph&\zeroelph&\zeroelph&\zeroelph&\zeroelph&\zeroelph&\zeroelph&\zeroelph&\zeroelph&\phantom{-}1&\zeroelph\\
\zeroelph&\zeroelph&\zeroelph&\zeroelph&\zeroelph&\zeroelph&\zeroelph&\zeroelph&\zeroelph&\zeroelph&\zeroelph&\phantom{-}1\\
\phantom{1}
\end{psmallmatrix}.
\end{displaymath}
This matrix is simple to compute, as each row $i$ contains a single $-1$ corresponding to the successor of the element $x_i$ in its chain. In other words, each row contains only the first element of the corresponding row in $V$ from \eqref{eq:zuv}. In our case, the first row of $V$ contains a $1$ in column $3$, therefore we have a $-1$ in the same position in $P^{-1}B^{-1}P$. 



\section{Algorithms}
\label{sec:alg}

In this section, we provide explicit algorithms to compute all the processes defined in \cref{sec:back,sec:main}. 
We start by describing an algorithm that computes a minimal chain decomposition of a DAG by reducing the task to a flow problem~\cite{fulkerson}. 
We then give an algorithm based on \cite{klaus-simon} that uses this chain decomposition to compute the $\niv$ map of the DAG. We finally use these two structures in the zeta and M\"obius transforms algorithms.

We conclude this section with a strategy for parallelization.

\mypar{Chain decomposition}
Our fast transform algorithms are based on the chain decomposition of a DAG. Such a decomposition is also known as vertex-disjoint path cover and different algorithms are available \cite{sparse-mpc,fulkerson,Hopcroft1973,klaus-simon}. By~\cref{th:dilworth}, the minimal chain decomposition contains $k = \width(\myP)$ chains. Since the number of operations of both zeta and M\"obius transform are bounded by $\n k - k$, we choose an algorithm that minimizes $k$.

We use the algorithm from~\cite{fulkerson} shown in~\cref{alg:chain-dec}. Given a DAG $G=(\myP,\E)$, it finds a minimal chain decomposition by finding a maximum matching (set of vertex-disjoint edges) of the bipartite graph  $G'=(\myP_1 \cup \myP_2,\E')$~\cite{fulkerson}. The vertex set of $G'$ is formed by duplicating each node in $\myP$ and taking the union of two sets $\myP_1$ and $\myP_2$, which contain the original vertices and the copies, respectively. We denote the copy of a vertex $v_1$ by $v_2$. The edge set $\E'$ is formed by all the pairs $(v_1,u_2) \in \myP_1 \times \myP_2$ such that $(v,u) \in \E$. In other words, we split each edge $(v,u) \in \E$ and link $v$ to the copy of $u$. We store the result of the maximum matching on $G$ in a auxiliary map we call $\next$, which corresponds to the function defined in~\cref{subsec:reach}.


The complexity of~\cref{alg:chain-dec} is determined by the maximum matching algorithm used. In our implementation, we use a \emph{push-relabel max flow} algorithm with a $\On(|V|^3)$ time complexity~\cite{max-flow}, which, however, is usually faster in practice.

\begin{algorithm}[t]
\caption{Chain Decomposition}
\label{alg:chain-dec}
\begin{algorithmic}[1]
\REQUIRE{A DAG $G=(\myP, \E)$.}
\ENSURE{A chain decomposition represented by the array $\next$.}
\STATE{Construct $\myP_1 \gets \{v_1 : v \in \myP\}$}
\STATE{Construct $\myP_2 \gets \{v_2 : v \in \myP\}$}
\STATE{Construct $\E' \gets \{ (v_1, u_2) : v_1,u_2 \in \myP_1 \cup \myP_2 \wedge (v,u) \in \E \}$}
\STATE{Initialize $G' \gets (\myP', \E')$}
\STATE{$\next \gets $ Maximum matching on $G'$} 
\RETURN $\next$
\end{algorithmic}
\end{algorithm}


\mypar{Niv computation}
Given a chain decomposition of a poset $\myP$ obtained via~\cref{alg:chain-dec}, we compute $\niv$ for $\myP$ by applying~\cref{thm:computenivj} to each vertex of the starting DAG. Note that in~\cref{thm:computenivj} we always have that $v \in \niv(v)$, therefore, in practice, we reduce memory usage by avoiding storing these self-loops. Here, we include them for simplicity and show in \cref{alg:niv-comp} the algorithm based on~\cite{klaus-simon} to compute $\niv$ for all vertices.

\begin{algorithm}[t]
\caption{Niv Computation}
\label{alg:niv-comp}
\begin{algorithmic}[1]
\REQUIRE{A DAG $G=(\myP, \E)$ and the $\id$ map of the chain decomposition.}
\ENSURE{$\niv$ of $G$.}
\STATE{$\niv_j \gets \{\infty, \dots, \infty \}$}  \COMMENT{Store $\niv_j$ as an array.}
\FOR{$v \gets \n$ to $1$}
\STATE{$\niv[v] \gets \emptyset$}
    \FOR{$w$ in $\adj(v)$} \label{l:firstloopniv}
        \IF{$w < \niv_j[\id[w]]$}
            \FOR{$p$ in $\niv[w]$}
                \STATE{$\niv_j[\id[p]] \gets \min(\niv_j[\id[p]],p)$}
            \ENDFOR
        \ENDIF
    \ENDFOR
    \FOR{$s \gets 1$ to $k$} \label{l:secondloopniv}
        \IF{$\niv_j[s] \neq \infty$} 
	        \STATE{$\niv[v] \gets \niv[v] \cup \{\niv_j[s]\}$}
        \ENDIF
        \STATE{$\niv_j[s] \gets \infty$}
    \ENDFOR
\ENDFOR
\RETURN $\niv$
\end{algorithmic}
\end{algorithm}

In the outer-most loop, we iterate over the elements of the poset. We construct $\niv$ for each vertex in two steps, executed by the two loops starting at lines~\ref{l:firstloopniv} and~\ref{l:secondloopniv}, respectively. The first loop finds all $\niv_j(v)$ values by iterating over the children of $v$. The second loop constructs $\niv(v)$ from $\niv_j$.

The complexity of this algorithm is shown~\cite{klaus-simon} to be $\On(|\E| + |\Ered|\cdot k)$, where $\Ered$ is the set of edges in the \emph{cover graph} of $\G$. The total memory requirement after~\cref{alg:chain-dec} and~\cref{alg:niv-comp} is $\n + \On(\n \cdot k)$, where $\n$ is due to the $\next$ array, and $\On(\n \cdot k)$ is given by $\niv$.

\mypar{Fast zeta transform}
We now provide the algorithm to compute~\cref{eq:zeta-fast} using the factorization $Z = UV$ obtained in~\cref{sec:main}. In particular, we compute the transform $UV \coord{x} = \coord{y}$ in two consecutive steps $V \coord{x} = \coord{x}'$ and $U \coord{x}' = \coord{y}$. However, we use the fact that the matrix are upper triangular to merge the two operations. Indeed, each $i$-entry of $\coord{x}'$, which we denote by $x_i'$, can be computed using $\next(x_i')$ as $x_i' = x_{\smash{\next(x_i')}}' + x_i$. To ensure that we have already computed $x_{\smash{\next(x_i')}}'$ we start the computation from the last row of the matrices. This, combined with the matrices being upper triangular, allows to compute $y_i = U x_i'$ immediately. This reasoning is translated into pseudocode in~\cref{alg:zeta}, where we use $\niv$ directly instead of the matrix $U$ to highlight the sparse computations we perform. Note that when considering a vertex $v$, which is the last element of a chain, the assignment at line~\ref{l:ass} is correct because by definition $\smash{\next(v)} = v$, and therefore $x_{\smash{\next(v)}}' = 0$.
\begin{algorithm}[t]
\caption{Fast Zeta Transform}
\label{alg:zeta}
\begin{algorithmic}[1]
\REQUIRE{Vector $\coord{x}$, $\next$ and $\niv$.}
\ENSURE{Transformed vector $\coord{y} = UV\coord{x}$.}
\STATE{$x' \gets \{0,\dots,0\}$}
\FOR{$i \gets \n$ to $1$}
    \STATE{$x'[i] \gets x'[\next[i]] + x[i]$} \label{l:ass}
    \FOR{$j$ in $\niv[i]$}
        \STATE{$y[i] \gets y[i] + x'[j]$}
    \ENDFOR
\ENDFOR
\end{algorithmic}
\end{algorithm}


As discussed in~\cref{sec:main}, the work required for~\cref{alg:zeta} is indeed $\nivsize - k \leq \n k - k$. Including the $k$ additions with zero in line~\ref{l:ass} yields exactly $\nivsize$ operations. No additional work is needed for the construction of the $\coord{x'}$ vector.

\mypar{Fast M\"obius transform}
Analogously, we now give an algorithm to compute the fast M\"obius transform we derived in~\cref{sec:main}, by solving the linear system $\coord{x} = Z \coord{y} = UV \coord{y}$. First, we solve the linear system $\coord{x} = U \coord{y}'$ via backsubstitution, then we compute the transform $V^{-1} \coord{y}' = \coord{y}$. These two steps can be seen at lines~\ref{l:firstloop} and~\ref{l:secondloop} of~\cref{alg:mob}. In contrast to the zeta transform, these two loops cannot be merged due to the data dependencies of backsubstitution. Additionally, we do not need to consider the vertex itself when looping over $\niv$ in line~\ref{l:innerloop}. An optimization we do in practice is to avoid storing these entries when computing $\niv$, and we change~\cref{alg:zeta,alg:mob} to not consider them.


\begin{algorithm}[t]
\caption{Fast M\"obius Transform}
\label{alg:mob}
\begin{algorithmic}[1]
\REQUIRE{Vector $\coord{x}$, $\next$ and $\niv$.}
\ENSURE{Transformed vector $\coord{y} = UV\coord{x}$.}
\STATE{$y' \gets \{0,\dots,0\}$}
\FOR{$i \gets \n$ to $1$} \label{l:firstloop}
    \STATE{$y'[i] \gets x[i]$}
    \FOR{$j$ in $\niv[i] \setminus i$} \label{l:innerloop}
        \STATE{$y'[i] \gets y'[i] - y'[j]$}
    \ENDFOR
\ENDFOR
\FOR{$i \gets \n$ to $1$} \label{l:secondloop}
    \IF{$\next(i) \neq i$}
    	\STATE{$y[i] \gets y'[i] - y'[\next[[i]]$}
	\ELSE
		\STATE{$y[i] \gets y'[i]$}
	\ENDIF
\ENDFOR
\end{algorithmic}
\end{algorithm}



As we derived in~\cref{sec:main}, the total work of~\cref{alg:mob} consists of $\nivsize - k \leq \n k - k$ operations.

\mypar{Parallel Implementation} 
As a final contribution, we propose a parallelization strategy to compute M\"obius and zeta transform with~\cref{alg:zeta,alg:mob}. The main idea is to find subsets of $\myP$ for which all vertices can be processed in parallel without data races. Since for both transforms the $i$-th coordinate of the output vector is constructed using only values indexed by $j \partleq i$ in $\myP$, we need to find subsets of pairwise non-comparable elements, i.e., antichains in $\myP$. Our solution thus uses a partition $\Lc = \{\Lc_1, \dots, \Lc_\ell\}$ of the poset $\myP$ where each subset $\Lc_i$ is an antichain of $\myP$ and the computation proceeds by executing one after the other in parallel. For a high degree of parallelism, these antichains should be large, and thus we use a minimal such partition whose length is known through the dual of~\cref{th:dilworth}:

\begin{theorem}[Mirsky's Antichain Decomposition Theorem~\cite{mirsky}]\label{th:mirsky} 
	We denote with $\ell = \length(\myP)$ the longest chain of the DAG $\G = (\myP, \E)$. Then, a minimal antichain decomposition of a DAG $\G$ has $\ell$ elements.
\end{theorem}

\begin{figure}
    \centering
    \includegraphics[scale=1.3]{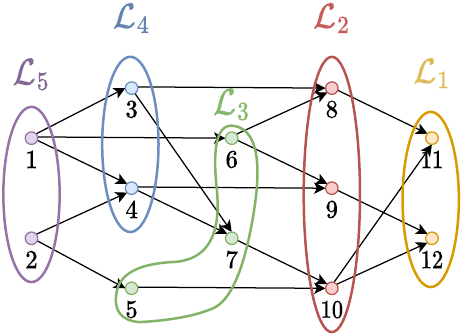}
    \caption{$\Lc$ of the example poset. Vertices of the same color belong to the same $\Lc_i$.}
    \label{fig:levels-ex}
\end{figure}

We give $\Lc$ for our example poset in~\cref{fig:levels-ex}. Since the longest chain in the poset contains $5$ vertices $\{1, 4, 7, 10, 12\}$ and we have $5$ subsets, the partition is minimal. Note that the minimal antichain decomposition is not unique. For example, we could choose $\Lc_4 = \{3, 4, 5\}$ and $\Lc_3 = \{6, 7\}$ as well. In our computations, this choice is only marginally relevant since the bottleneck is the synchronization overhead is given by the transition between one antichain and the next. Consequently, we only require $\Lc$ to be minimal. The final algorithm is sketched in~\cref{alg:scheduling}. As mentioned, the partition $\Lc$ allows the computation of all coordinates indexed by $v \in \Lc_i$ of the transformed signal in parallel without locking. Thus, we only need a barrier at line~(\ref{ln:barrier}) to synchronize the threads between antichains explicitly. Similar ideas are used in~\cite{Gupta-Kumar, Naumov2011ParallelSO, sparse-parallel-solve} to improve parallel triangular-solve algorithms.

\begin{algorithm}[t]
\caption{Scheduling strategy}
\label{alg:scheduling}
\begin{algorithmic}[1]
\REQUIRE{A partition $\Lc = \{ \Lc_1, \dots, \Lc_\ell \}$.}
\ENSURE{A transformed signal $\coord{y}$}
\FORALL{$\Lc_i$ in $\Lc$}
    \FORALL{$v$ in $\Lc_i$}
		\STATE{Compute $y_v$ on available thread}
	\ENDFOR
    \STATE{barrier()} \label{ln:barrier}
\ENDFOR
\end{algorithmic}
\end{algorithm}


\begin{algorithm}[t]
\caption{Antichain Decomposition}
\label{alg:partition}
\begin{algorithmic}[1]
\REQUIRE{A DAG $G = (\myP, \E)$.}
\ENSURE{A partition $\Lc$}
\STATE{Construct $G'=(\myP,\E'=\{(u,v):u,v \in \myP \wedge (v,u)\in\E\}$)}
\STATE{$\delta \gets \{\text{array mapping each vertex to its out-degree}\}$}
\STATE{$\Lc_1 \gets \{ \text{leaves of } \G \}$}
\STATE{$i \gets 1$}
\WHILE{$\Lc_i \neq \emptyset$}
    \FORALL{$v$ in $\Lc_i$}
        \FORALL{$w$ in $\adj(v)$ in $G'$}
            \STATE{$\delta[w] \gets \delta[w] - 1$}
            \IF{$\delta[w] = 0$}
                \STATE{$\Lc_{i+1} \gets \Lc_{i+1} \cup \{w\}$}
            \ENDIF
        \ENDFOR
    \ENDFOR
    \STATE{$i \gets i + 1$}
\ENDWHILE
\end{algorithmic}
\end{algorithm}

We compute a minimal partition into antichains using~\cref{alg:partition}. The main idea is to traverse the DAG in reverse order and, at each step $i$, remove the leaves of the DAG and insert them in $\Lc_i$. To do so, we use an auxiliary data structure that keeps track of the number of children of each node. Whenever one of these counters is $0$, we add the corresponding node to the subset $\Lc_i$ and decrement the parents' counter.




We theoretically evaluate our parallelization using the PRAM model~\cite{PRAM}, which is an extension of the classic RAM model that includes parallel instructions. The main component of a PRAM model is the \emph{computation DAG} that summarizes data dependencies among the operations issued during the execution. Each node is an operation, and the relevant characteristics of this DAG are its \emph{work} $W(n)$ defined as the number of nodes, and \emph{depth} $D(n)$ defined as the length of the DAG. The parameter $n$ is the size of the input. Assuming no synchronization overhead and an infinite number of processors, the theoretical speedup in the PRAM model is then $W(n) / D(n)$, which is also called \emph{average parallelism}. 

In the case of our transforms, the work is given in both cases by the number of operations we found in~\cref{sec:main}. This means that $W_Z(n) = W_M(n) =  \nivsize - k$ for both computations. On the other hand, the depth varies. We can find an upper bound to the depth by considering the matrix view given in~\eqref{eq:zuv}. There, the depth of the multiplication $\coord{c} = V \coord{x}$  is given by the size of the longest chain in our chain decomposition, which we denote by $\q$, as the reduction must be sequential. For $\coord{y} = U \coord{x}$ instead, we can execute all dot products in parallel. Hence, we have a depth of $\log(k)$ as we can perform a tree reduction for each row with at most $k$ elements. From~~\cref{alg:zeta} we see that the multiplication $V \coord{x}$ is done implicitly, the total depth for the zeta transform is, therefore, bounded as $D_Z(n) \leq \q + \log(k)$. Since we perform our M\"obius transform as a sparse triangular solve, the depth of the PRAM computational DAG depends on the topology of $\myP$. First, we have that $\coord{y} = P^{-1} V^{-1} P \coord{y}'$ has depth $1$, since it corresponds to $\n - k$ subtractions that can be done in parallel. For the triangular solve itself, we have that for each layer $\Lc_i$ in $\Lc$ we perform at most $k$ operations. The depth is then $D_M(n) \leq \ell k + 1$. With these two results, we can lower bound the average parallelism for the zeta and M\"obius transforms, respectively, by
\begin{displaymath}
	\frac{W_Z(\n)}{D_Z(\n)} \geq \frac{\nivsize - k}{\q + \log(k)}
	\quad\text{ and }\quad 
	\frac{W_M(\n)}{D_M(\n)} \geq \frac{\nivsize - k}{\ell k + 1}.
\end{displaymath}



\section{Experimental results} \label{sec:experiments}

We present experimental results of our zeta and M\"obius transform algorithms to demonstrate actual achievable runtimes and scalability with input size and number of threads. In particular, we show that we can compute the transforms on DAGs with millions of modes in seconds on a standard workstation.

All algorithms are implemented in C\texttt{++} and compiled with gcc 9.3.0. The computations are performed on an AMD EPYC 7742 64 core CPU with 504 GB of main memory.

\mypar{Random input DAGs}
To obtain valid estimates, we repeatedly measure our results on random DAGs of increasing size and different densities. To generate the input DAGs we use the Erd\H{o}s-Renyi~\cite{erdos} model, in which each edge $(v,u)$ is generated with probability $p$ but only the direction with $v \partl u$ is kept to make the adjacency matrix upper triangular. By increasing $p$ we thus increase the density, i.e., the average degree of each node. We choose $p$ to obtain DAGs with an average degree $\delta$ per node of $4$ and $6$. 



\begin{figure}
	\centering
	\includegraphics[scale=0.2995]{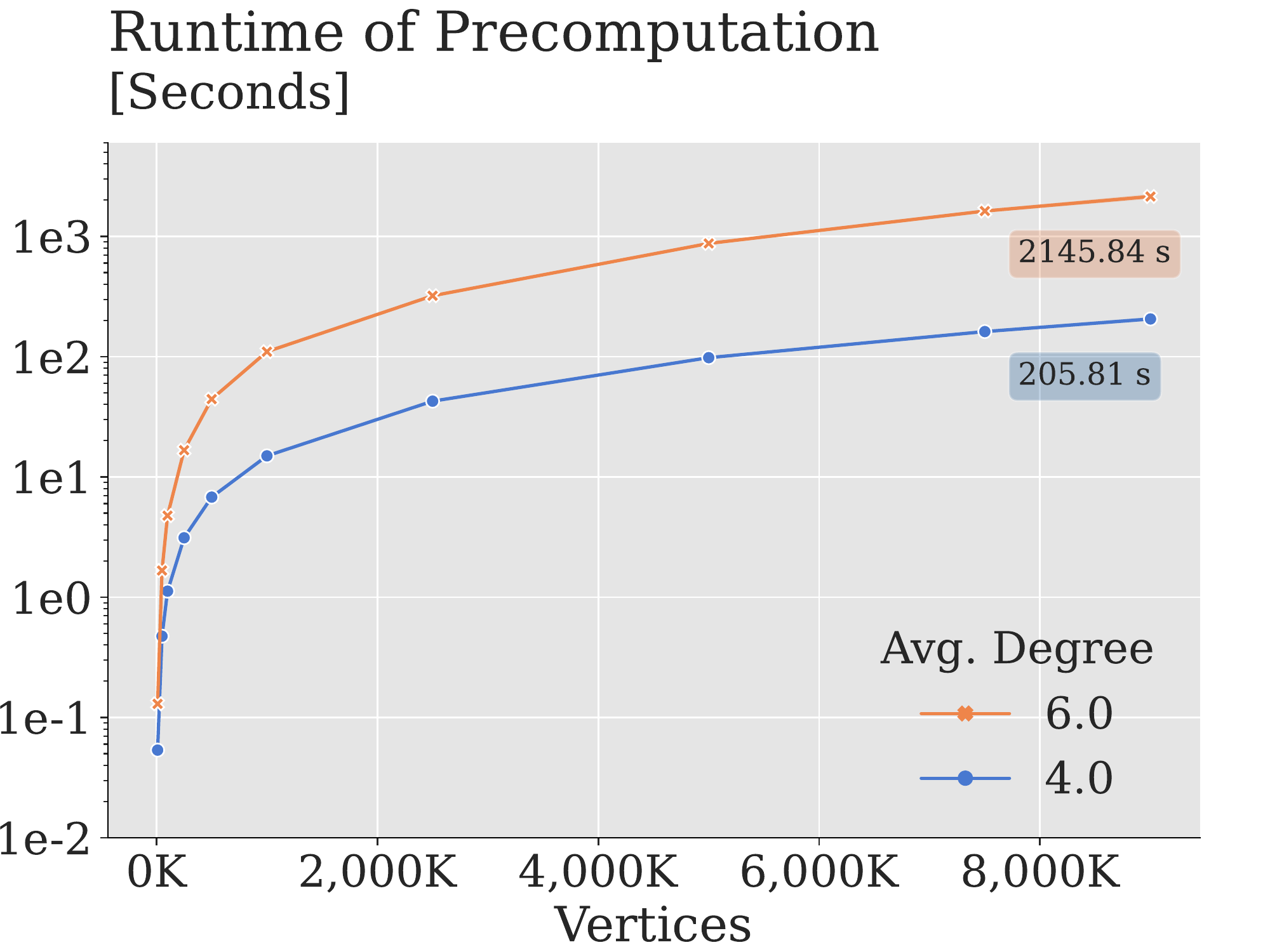}
	\caption{Cumulative runtime in seconds measured for the needed precomputation in~\cref{sec:alg}.}\label{fig:runtime}
\end{figure}

\mypar{Runtime of precomputation} \cref{fig:runtime} shows the runtime of the needed precomputation, i.e., the chain decomposition and niv computation of \cref{alg:chain-dec,alg:niv-comp}, respectively. This computation is only done once per DAG; however, it still has to be fast enough to amortize with a reasonable number of subsequent computations of M\"obius or zeta transforms.

The plot shows the runtime for 15 random DAGs for both average degrees (4 and 6) for a number of nodes ranging from 
$10$k up to $9$M. Note that the plot (and also the subsequent plots) also contains the $95\%$ confidence intervals, which here are too narrow to be seen. We observe that for the largest size and average degree 4, the computation takes about 3 minutes, while for average degree 6 it is already about 36 minutes. This trend is expected as a more edges correspond to significantly more possible paths that must be checked when constructing $\niv$ using~\cref{thm:computenivj}. In summary, this shows that large DAGs can be preprocessed provided they are sufficiently sparse.

\begin{figure}
	\centering
	\begin{subfigure}[b]{0.49\textwidth}\label{subfig:runtime-zeta}
	\centering
		\includegraphics[scale=0.2995]{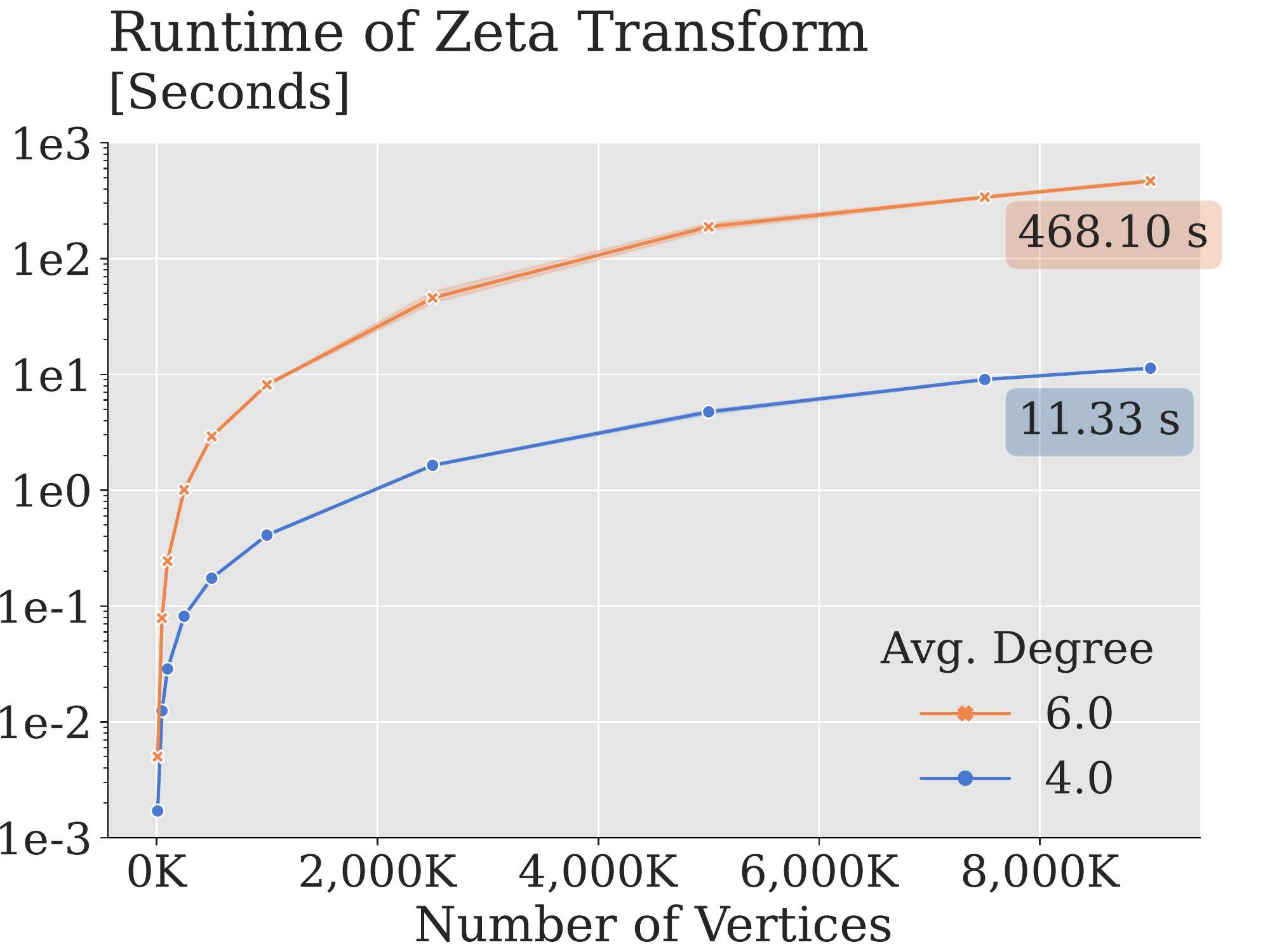}
		\caption{}
	\end{subfigure}
	\begin{subfigure}[b]{0.49\textwidth}\label{subfig:runtime-mob}
	\centering
		\includegraphics[scale=0.2995]{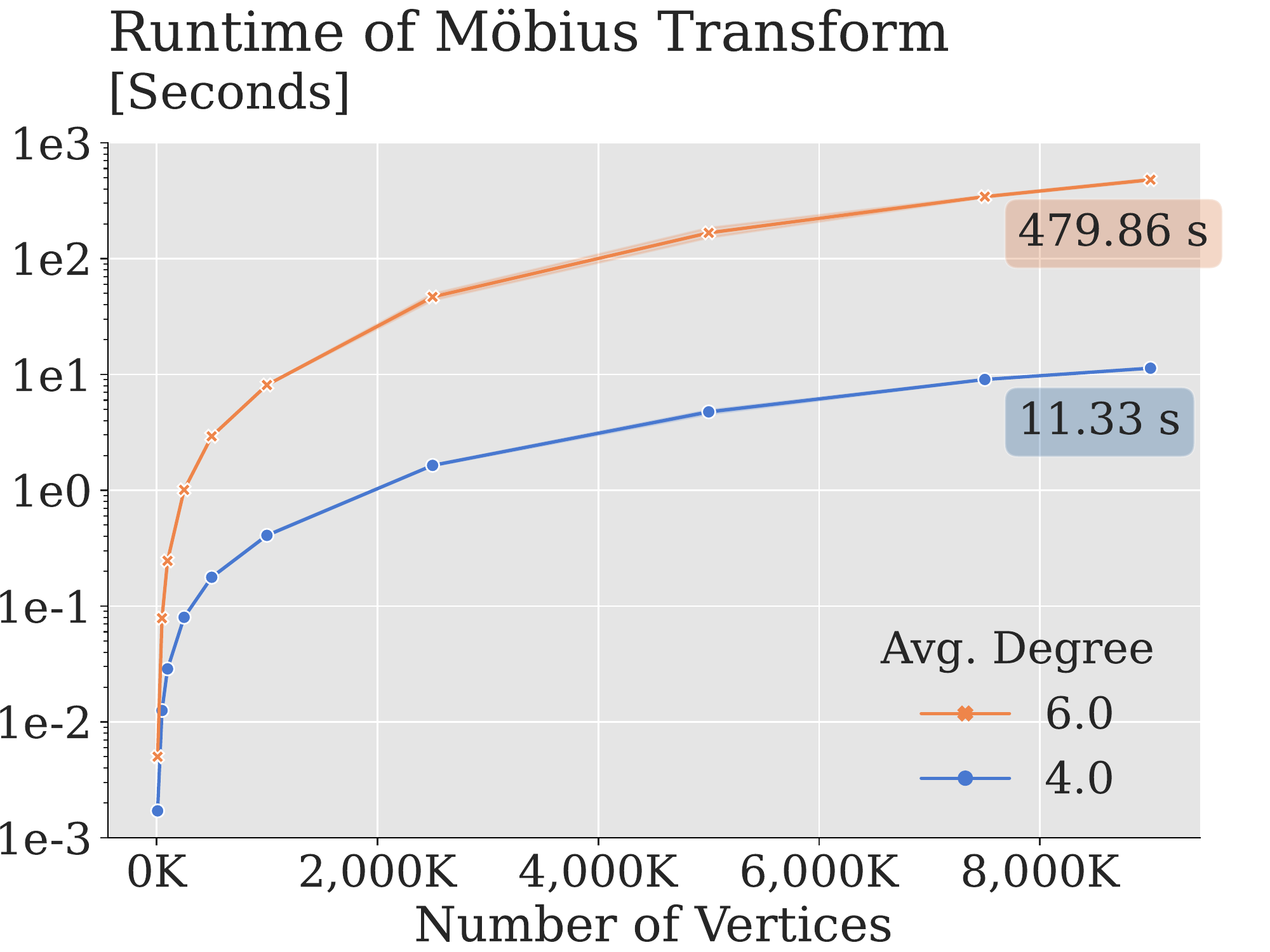}
		\caption{}
	\end{subfigure}
	\caption{Runtime in seconds of zeta transform (\cref{alg:zeta}) and M\"obius transform (\cref{alg:mob}).}\label{fig:runtime-trans}
\end{figure}

\mypar{Runtime of zeta and M\"obius transforms}
\cref{fig:runtime-trans} shows the runtime of zeta and M\"obius transform, again for the average degrees 4 and 6 and for the same range of number of nodes as in \cref{fig:runtime}. Both take about the same time which corresponds to the bounds in~\cref{sec:main}, and indeed, the implementations have the same number of floating point operations. The runtimes are about 19 (degree 4) and 4 times (degree 6) faster than the precomputation, which also confirms its fast amortization.

\begin{figure}
	\centering
	\begin{subfigure}[b]{0.49\textwidth}\label{subfig:speedup-zeta}
		\centering
		\includegraphics[scale=0.3145]{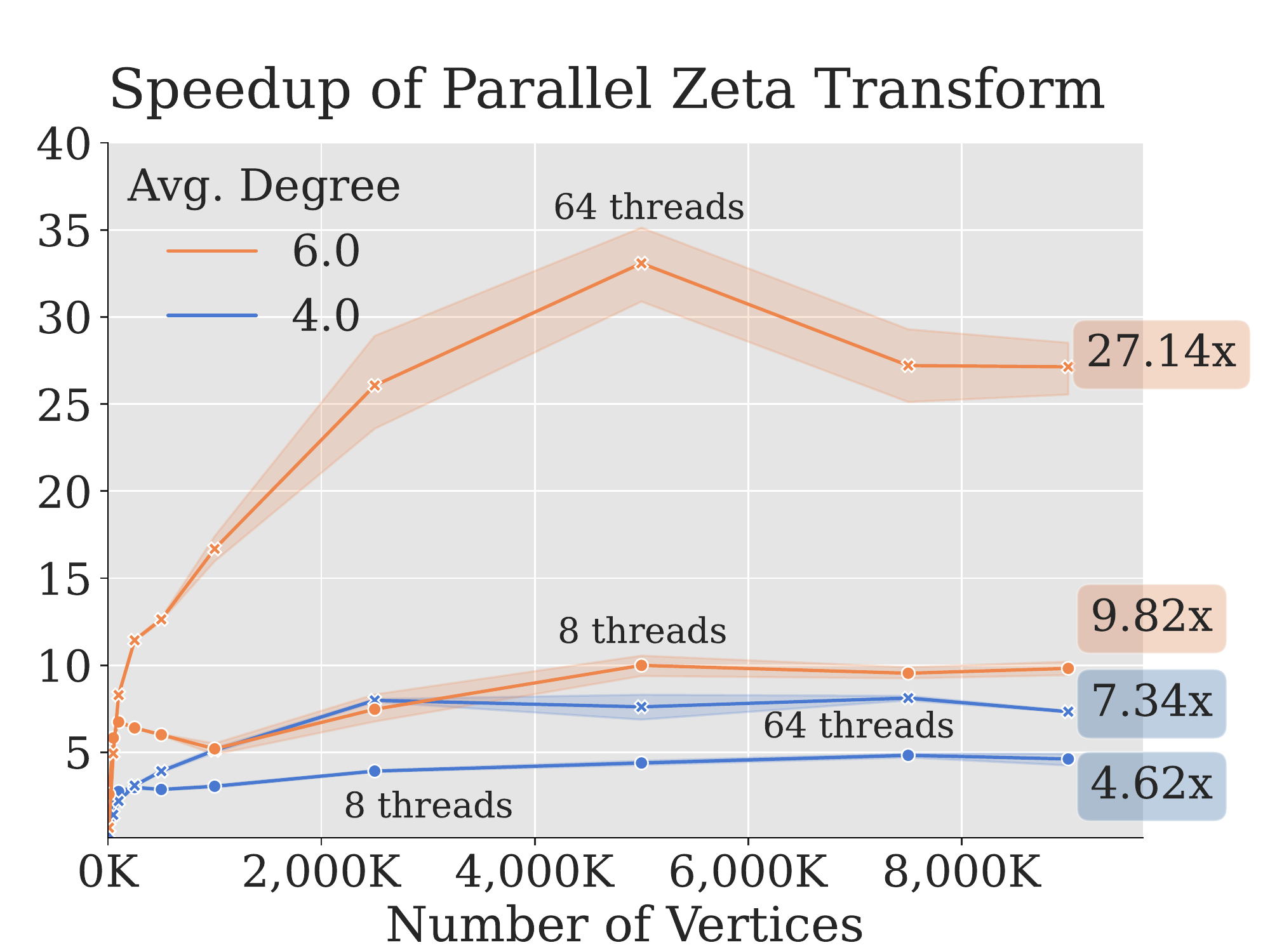}
		\caption{}
	\end{subfigure}
	\begin{subfigure}[b]{0.49\textwidth}\label{subfig:speedup-mob}
		\centering
		\includegraphics[scale=0.3145]{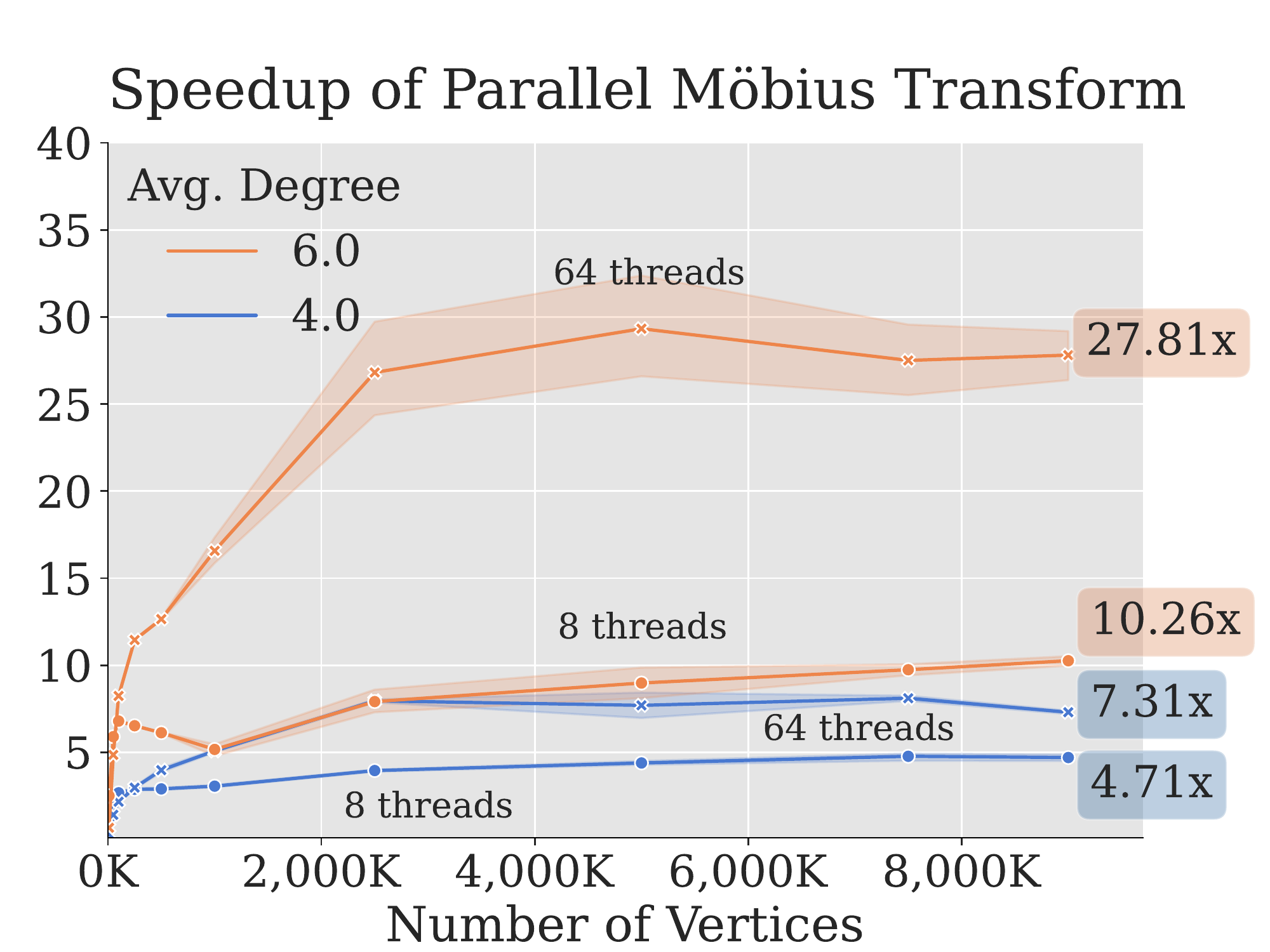}
		\caption{}
	\end{subfigure}
	\caption{Parallel speedup using $8$ and $64$ threads obtained with our parallel implementations based on~\cref{alg:zeta,alg:mob}.}\label{fig:speedup-par}
\end{figure}

\mypar{Parallel speedup}
The last measurements we present show parallel speedup we obtain using the scheduling strategy in~\cref{alg:scheduling}. In particular, we plot in~\cref{fig:speedup-par} the parallel speedup obtained over 1 core when using $8$ and $64$ cores, respectively, again as a function of the number of nodes. The speedup is significant in all cases, generally higher for larger and denser DAGs. For 8 threads on the denser DAG, the speedup is even superlinear, possibly due to the larger available cache size. For the largest DAGs with 9M nodes, the speedup with all 64 cores is about $7\times$ for average degree 4 and $27\times$ for average degree 6, i.e., the latter runs in about only 17 seconds.

In summary, our work makes both zeta and M\"obius transform practical and efficient for very large DAGs with millions of nodes, provided they are sufficiently sparse.

\section{Conclusions}

We presented a novel approach to the fast computation of the zeta and M\"obius transforms on arbitrary posets or DAGs. The key idea was to use the chain decomposition of a poset $\myP$ and the $\niv$ concept induced by this decomposition to factor the transform matrices into a product of two, a form of common subexpression elimination. Our approach reduces both memory and time complexity from $\On(\n^2)$ to $\On(\n k)$, where $\n$ is the number of elements in the poset, or nodes in the DAG, and $k$ is its width. 
The results with an actual implementation show that our algorithms for both transforms scale to DAGs with millions of nodes for sparse graphs and offer excellent parallel speedup on a current work station.

\bibliographystyle{unsrt}
\bibliography{references}

\begin{thebibliography}{10}

\bibitem{rota}
G.~C. Rota.
\newblock On the foundations of combinatorial theory i. theory of möbius
  functions, 1964.

\bibitem{stanley_2011}
Richard~P. Stanley.
\newblock {\em Enumerative Combinatorics}, volume~1 of {\em Cambridge Studies
  in Advanced Mathematics}.
\newblock Cambridge University Press, 2 edition, 2011.

\bibitem{fulkerson}
D.~R. Fulkerson.
\newblock Note on dilworth's decomposition theorem for partially ordered sets.
\newblock {\em Proceedings of the American Mathematical Society},
  7(4):701--702, 1956.

\bibitem{klaus-simon}
K.~Simon.
\newblock An improved algorithm for transitive closure on acyclic digraphs.,
  1986.

\bibitem{Graetzer:11}
G.~Gr{\"a}tzer.
\newblock {\em Lattice Theory: Foundation}.
\newblock Birkh{\"a}user, 2011.

\bibitem{ABDALI1985257}
S.~K. Abdali and B.~D. Saunders.
\newblock Transitive closure and related semiring properties via eliminants.
\newblock {\em Theoretical Computer Science}, 40:257--274, 1985.
\newblock Eleventh International Colloquium on Automata, Languages and
  Programming.

\bibitem{LEHMANN197759}
D.~J. Lehmann.
\newblock Algebraic structures for transitive closure.
\newblock {\em Theoretical Computer Science}, 4(1):59--76, 1977.

\bibitem{yates1937design}
F.~Yates.
\newblock The design and analysis of factorial experiments.
\newblock 1937.

\bibitem{kennes-moebius}
R.~Kennes and P.~Smets.
\newblock Computational aspects of the mobius transform, 2013.

\bibitem{moebius-computational-aspects}
R.~Kennes.
\newblock Computational aspects of the mobius transformation of graphs.
\newblock {\em IEEE Transactions on Systems, Man, and Cybernetics},
  22(2):201--223, 1992.

\bibitem{few-irreducibles}
A.~Bj\"{o}rklund, T.~Husfeldt, P.~Kaski, M.~Koivisto, J.~Nederlof, and
  P.~Parviainen.
\newblock Fast zeta transforms for lattices with few irreducibles.
\newblock {\em ACM Trans. Algorithms}, 12(1), 2016.

\bibitem{kaski2016fast}
P.~Kaski, J.~Kohonen, and T.~Westerb{\"a}ck.
\newblock {Fast M{\"o}bius Inversion in Semimodular Lattices and ER-labelable
  Posets}.
\newblock {\em The Electronic Journal of Combinatorics}, 23(3):P3.26, 2016.

\bibitem{valiant1986negation}
L.~G. Valiant.
\newblock {Negation is powerless for Boolean slice functions}.
\newblock {\em SIAM J.~Comp.}, 15(2):531--535, 1986.

\bibitem{schapire1996learning}
R.~E. Schapire and L.~M. Sellie.
\newblock Learning sparse multivariate polynomials over a field with queries
  and counterexamples.
\newblock {\em Journal of Computer and System Sciences}, 52(2):201--213, 1996.

\bibitem{malandro2010fast}
M.~Malandro and D.~Rockmore.
\newblock Fast {F}ourier transforms for the rook monoid.
\newblock {\em Trans Am Math Soc}, 362(2):1009--1045, 2010.

\bibitem{malandro2015fourier}
M.~E. Malandro.
\newblock Fourier inversion for finite inverse semigroups.
\newblock {\em SIAM J.~Disc.~Math.}, 29(1):269--296, 2015.

\bibitem{maslen2018efficient}
D.~Maslen, D.~N. Rockmore, and S.~Wolff.
\newblock The efficient computation of {F}ourier transforms on semisimple
  algebras.
\newblock {\em J.~Fourier Anal.~Appl.}, 24(5):1377--1400, 2018.

\bibitem{mob_appl_combinatorics}
E.~A. Bender and J.~R. Goldman.
\newblock On the applications of möbius inversion in combinatorial analysis.
\newblock {\em The American Mathematical Monthly}, 82(8):789--803, 1975.

\bibitem{ordered-sets}
I.~Rival.
\newblock {\em {Ordered Sets}}.
\newblock 1982.

\bibitem{game_theory}
M.~Grabisch.
\newblock {\em Set Functions, Games and Capacities in Decision Making}.
\newblock Springer Publishing Company, Incorporated, 1st edition, 2016.

\bibitem{dempster-shafer1}
A.~P. Dempster.
\newblock {Upper and Lower Probabilities Induced by a Multivalued Mapping}.
\newblock {\em The Annals of Mathematical Statistics}, 38(2):325 -- 339, 1967.

\bibitem{dempster-shafer0}
G.~Shafer.
\newblock {\em A Mathematical Theory of Evidence}.
\newblock Princeton University Press, 1976.

\bibitem{koivisto-bayes-net}
M.~Koivisto and K.~Sood.
\newblock Exact bayesian structure discovery in bayesian networks.
\newblock {\em J. Mach. Learn. Res.}, 5:549–573, dec 2004.

\bibitem{p_databases_paper}
N.~Dalvi, K.~Schnaitter, and D.~Suciu.
\newblock Computing query probability with incidence algebras.
\newblock In {\em Proceedings of the Twenty-Ninth ACM SIGMOD-SIGACT-SIGART
  Symposium on Principles of Database Systems}, page 203–214. Association for
  Computing Machinery, 2010.

\bibitem{p_database}
N.~Dalvi and D.~Suciu.
\newblock The dichotomy of probabilistic inference for unions of conjunctive
  queries.
\newblock 59(6), 2013.

\bibitem{poset-decentralized-control}
P.~Shah and P.~A. Parrilo.
\newblock A partial order approach to decentralized control.
\newblock In {\em 2008 47th IEEE Conference on Decision and Control}, pages
  4351--4356, 2008.

\bibitem{moebius-controller}
P.~Shah and P.~A. Parrilo.
\newblock An optimal controller architecture for poset-causal systems.
\newblock In {\em 2011 50th IEEE Conference on Decision and Control and
  European Control Conference}, pages 5522--5528, 2011.

\bibitem{cluster_expansion}
D.~Babić and N.~Trinajstić.
\newblock Möbius inversion on a poset of a graph and its acyclic subgraphs.
\newblock {\em Discrete Applied Mathematics}, 67(1):5--11, 1996.

\bibitem{molecule}
B.~Schweinhart, D.~Rodney, and J.~K. Mason.
\newblock Statistical topology of bond networks with applications to silica.
\newblock {\em Phys. Rev. E}, 101, May 2020.

\bibitem{lattice}
M.~Püschel.
\newblock A discrete signal processing framework for meet/join lattices with
  applications to hypergraphs and trees.
\newblock In {\em ICASSP 2019 - 2019 IEEE International Conference on
  Acoustics, Speech and Signal Processing (ICASSP)}, pages 5371--5375, 2019.

\bibitem{discrete_meet_join_lattices}
M.~Püschel, B.~Seifert, and C.~Wendler.
\newblock Discrete signal processing on meet/join lattices.
\newblock {\em IEEE Transactions on Signal Processing}, 69:3571–3584, 2021.

\bibitem{seifert2022learning}
Bastian Seifert, Chris Wendler, and Markus P{\"u}schel.
\newblock Learning fourier-sparse functions on dags.
\newblock In {\em ICLR2022 Workshop on the Elements of Reasoning: Objects,
  Structure and Causality}, 2022.

\bibitem{Seifert22}
B.~Seifert, C.~Wendler, and M.~Püschel.
\newblock Causal fourier analysis on directed acyclic graphs and posets, 2022.

\bibitem{welsh2010matroid}
D.J.A. Welsh.
\newblock {\em Matroid Theory}.
\newblock Dover Publications, 2010.

\bibitem{dilworth}
R.~P. Dilworth.
\newblock A decomposition theorem for partially ordered sets.
\newblock {\em Annals of Mathematics}, 51(1):161--166, 1950.

\bibitem{max-flow}
A.~V. Goldberg and R.~E. Tarjan.
\newblock A new approach to the maximum-flow problem.
\newblock {\em J. ACM}, 35(4):921–940, oct 1988.

\bibitem{sparse-mpc}
M.~Cáceres, M.~Cairo, B.~Mumey, R.~Rizzi, and A.~I. Tomescu.
\newblock {\em Sparsifying, Shrinking and Splicing for Minimum Path Cover in
  Parameterized Linear Time}, pages 359--376.

\bibitem{Hopcroft1973}
J.~E. Hopcroft and R.~M. Karp.
\newblock An n5/2 algorithm for maximum matchings in bipartite graphs.
\newblock {\em SIAM J. Comput.}, 2:225--231, 1973.

\bibitem{mirsky}
L.~Mirsky.
\newblock A dual of dilworth's decomposition theorem.
\newblock {\em The American Mathematical Monthly}, 78(8):876--877, 1971.

\bibitem{Gupta-Kumar}
A.~Gupta and V.~Kumar.
\newblock Parallel algorithms for forward and back substitution in direct
  solution of sparse linear systems.
\newblock Supercomputing '95, page 74–es, New York, NY, USA, 1995.
  Association for Computing Machinery.

\bibitem{Naumov2011ParallelSO}
M.~Naumov.
\newblock Parallel solution of sparse triangular linear systems in the
  preconditioned iterative methods on the gpu.
\newblock 2011.

\bibitem{sparse-parallel-solve}
J.~Park, M.~Smelyanskiy, N.~Sundaram, and P.~Dubey.
\newblock Sparsifying synchronization for high-performance shared-memory sparse
  triangular solver.
\newblock In {\em Supercomputing}, pages 124--140, Cham, 2014. Springer
  International Publishing.

\bibitem{PRAM}
J.~F. JaJa.
\newblock {\em PRAM (Parallel Random Access Machines)}, pages 1608--1615.
\newblock Springer US, Boston, MA, 2011.

\bibitem{erdos}
P.~Erd\H{o}s and A.~R\'{e}nyi.
\newblock On random graphs i.
\newblock {\em Publicationes Mathematicae Debrecen}, 6:290, 1959.

\end{thebibliography}
\end{document}